\documentclass[a4paper,10pt,oneside,onecolumn,number,preprint]{elsarticle} 
\biboptions{round,authoryear}
\usepackage{ecrc}
\volume{00}              
\firstpage{1}            
\journalname{ICARUS}     
\runauth{Carruba et al.} 
\jnltitlelogo{\bf ICARUS}    
\CopyrightLine{2011}{Elsevier Ltd. All rights reserved}
\usepackage{amssymb}
\usepackage{amsthm}
\usepackage{lineno}      
\usepackage[figuresright]{rotating}   
\usepackage[table]{xcolor}
\usepackage{caption} 
\usepackage{multicol, blindtext}
\definecolor{grey80}{rgb}{0.90,0.90,0.90}

\usepackage[dvipdfm]{hyperref}
\hypersetup{
   colorlinks=true,             
   linkcolor=blue,              
   urlcolor=blue,               
   anchorcolor=blue,            
   citecolor=blue,              
}
\begin{document}
\begin{frontmatter}
\title{\textcolor{white}{.}\\ \textcolor{white}{.}\\ \textcolor{white}{.}\\ \centering Characterizing the original ejection velocity field of the Koronis family} 
\author[1,2]{V. Carruba \corref{Corresponding}}\ead{vcarruba@feg.unesp.br}
\cortext[Corresponding]{Corresponding author}
\author[2]{D. Nesvorn\'{y}}
\author[1]{S. Aljbaae}
\address[1]{Univ. Estadual Paulista - UNESP, Grupo de Din\^amica Orbital \& Planetologia\,, Guaratinguet\'a\,, CEP 12516-410\,, SP\,, Brazil}
\address[2]{Department of Space Studies, Southwest Space Research Institute, 
1050 Walnut St.,Boulder, CO 80302, USA}
\begin{abstract}
An asteroid family forms as a result of a collision between an impactor and 
a parent body.  The fragments with ejection speeds higher than the escape 
velocity from the parent body can escape its gravitational pull.  The cloud of 
escaping debris can be identified by the proximity of orbits in proper 
element, or frequency, domains.  Obtaining estimates of the original 
ejection speed can provide valuable constraints on the physical 
processes occurring during collision, and used to calibrate impact 
simulations.  Unfortunately, proper elements of asteroids families
are modified by gravitational and non-gravitational effects, such as 
resonant dynamics, encounters with massive bodies, and the Yarkovsky
effect, such that information on the original ejection speeds is 
often lost, especially for older, more evolved families.

It has been recently suggested that the distribution in proper inclination
of the Koronis family may have not been significantly perturbed by local 
dynamics, and that information on the component of the ejection velocity 
that is perpendicular to the orbital plane ($v_W$), may still be 
available, at least in part.  In this work we estimate the magnitude 
of the original ejection velocity speeds of Koronis members using 
the observed distribution in proper eccentricity and inclination, and 
accounting for the spread caused by dynamical effects.  Our results show 
that i) the spread in the original ejection speeds is, to within a 15\%
error, inversely proportional to the fragment size, and ii) the minimum 
ejection velocity is of the order of 50 m/s, with larger values possible 
depending on the orbital configuration at the break-up.
\end{abstract}
\end{frontmatter}
\section{Introduction}\label{Introduction}
Asteroid families are the outcome of high-velocity collisions between 
asteroids.  If the fragments of collisions have speeds exceeding the 
escape velocity from the parent body, they may be ejected and form a 
swarm around the main body (or, in case of a catastrophic collision, around the 
family barycenter).  Since the terminal ejection speeds
are only a fraction of the orbital speed of most main belt bodies, the 
fragments are not ejected far and may be identified by the fact that 
they form clusters in the domain of proper elements $(a,e,sin(i))$ (or 
proper frequencies $(n,g,g+s)$ (see \citet{Knezevic_2000} for 
a description of the methods used to obtain proper elements and frequencies), 
with $a$, $e$ and $i$ being the asteroids 
proper semi-major axis, eccentricity, and inclination, and $n$, $g$, $s$ 
being the proper mean-motion, frequency of precession of the longitudes 
of pericenter, and of the node, respectively
\citep{Bendjoya_2002, Carruba_2007}.

Theoretically, ejection speeds of asteroid family members could be estimated 
from the distribution in proper elements $(a,e,sin(i))$ using Gauss'equations
\citep{Zappala_1996}, provided that both the true anomaly and the 
argument of perihelion of the family parent body are known (or estimated).  
In practice, several gravitational and non-gravitational effects, such
as resonant dynamics \citep{Morbidelli_1999}, close encounters 
with massive asteroids \citep{Carruba_2003}, and the Yarkovsky effect 
\citep{Bottke_2001} can change proper elements of asteroid families, so 
that information on the original ejection speeds is sometimes
lost, especially for older, more dynamically evolved groups.

Recently, however, \citet{Carruba_2015c}, based on arguments on the
current shape of the distribution in $\sin(i)$ of the Koronis family, 
suggested that information on the component of the ejection velocity that is 
perpendicular to the orbital plane ($v_W$) may still be available 
for this family, at least in part~\footnote{Information on the original
ejection velocity field could be in principle also obtainable from very
young families proper $e$ and $i$ distributions, such as for instance the 
Datura and Lorre clusters.  Here however we focus our attention on a 
relatively old family, to check whatever information may still available for 
more evolved groups.}.  In this work we extend the analysis 
of the previous paper, and try for the first time to obtain estimates
of the original spread of the ejection velocity field for the Koronis
group, based on the {\it eccentricity and inclination} distributions, 
rather than on the semi-major axis one, as previously attempted 
\citep{Carruba_2015b}.  By performing long-term simulations of fictitious 
Koronis members, we were able to evaluate the effect that dynamics may have had 
on the inclination distribution, and to estimate what 
the original distribution should have been, so allowing us to obtain a 
value of the spread in ejection speeds.
 
This paper is so divided: in Sect.~\ref{sec: fam_ide} we identify the Koronis
family and its halo, and we verify what is the current dispersion of the
family in proper $e$ and $\sin{(i)}$.  In Sect.~\ref{sec: loc_dyn} we 
study the effect that the local dynamics may have had on the orbital
evolution of Koronis family members, and determine to which extent the
original distribution in $e$ and $\sin{(i)}$ have been preserved.  In 
Sect.~\ref{sec: ej_field} we estimate what was the original dispersion
of the Koronis family in proper inclination, determine how the spread
in original inclination depended on the family members sizes, and estimate
what was the magnitude of the family original ejection velocity field.
Finally, in Sect.~\ref{sec: conc} we present our conclusions.

\section{Family identification}
\label{sec: fam_ide}

The Koronis family is located in a relatively dynamically quiet region,
where, with the exception of the $3{\nu}_6-2{\nu}_5 = g+2g_5-3g_6$ secular
resonance, a pericenter resonance that mostly modifies asteroid eccentricities,
no other major mean-motion or secular resonance exists 
\citep{Bottke_2001, Carruba_2013, Nesvorny_2015}. The long-term dynamics of 
the Koronis family should therefore had had a minor influence in the 
spreading in proper $i$ of members of the family, as we will also further 
investigate later on in this paper.  Despite being a relatively old 
family (2.65 Gyr at most if one use standard values for the parameters
describing the Yarkovsky force \citep{Broz_2013}, see \citet{Carruba_2015b} 
and references therein for more detail on this age estimate),
the current dispersion in proper $i$ of Koronis family members
could therefore still contain traces of information on the original ejection
speeds \citep{Cellino_2004, Cellino_2009}.

To investigate this hypothesis, we first obtained an estimate of the 
current family members.   We used data published in \citet{Nesvorny_2015}
for the Koronis family using the Hierarchical Clustering Method 
\citep{Zappala_1990}, and a 
velocity cutoff of 45 m/s in the $(a,e,sin(i))$ domain (see Table 2 
in \citet{Nesvorny_2015} for further details).  Since there
are two subfamilies inside the Koronis group, the Karin cluster 
(\citet{Nesvorny_2002, Nesvorny_2006}) and the Koronis 2 family 
\citep{Molnar_2009}, we eliminated these two subgroups from the list of 
Koronis members. This left us with a sample of 5163 objects, out of the 
original 5949 members of the whole Koronis and subfamilies sample.

To eliminate possible interlopers, we used the classification method of 
\citet{DeMeo_2013} that employs Sloan 
Digital Sky Survey-Moving Object Catalog data, fourth release (SDSS-MOC4 
hereafter, \citet{Ivezic_2001} to compute $gri$ slope and $z' -i'$ colors, 
and data from the three major photometric/spectroscopic
surveys: ECAS, Eight-Color Asteroid Analysis, 
\citet{Zellner_1985, Tholen_1989}, SMASS, Small Main Belt Spectroscopic 
Survey, \citet{Xu_1995, Bus_2002a, Bus_2002b}, and S3OS2, Small Solar 
System Objects Spectroscopic Survey, \citet{Lazzaro_2004}.
We obtained 896 observations in the SDSS-MOC4 catalog, and taxonomical 
information for 507 individual asteroids in the revised Koronis family 
(i.e., the Koronis family, less the two sub-families of Karin and
Koronis 2).  We found 3 C-type, 8 D-type, 29 X-type, 109 L-type, 259 
S-type, 98 K-type, and 1 A-type object, respectively.

Our results confirm the analysis of \citet{Binzel_1993, Carruba_2013, 
Carruba_2015b}: the Koronis family is an S-complex group, with a small 
percentage (40 objects, 7.9\% of the available sample) of C-complex 
interlopers.   This is further confirmed by the values of geometric albedos 
$p_V$ from the WISE mission \citep{Masiero_2012} for local asteroids.   Out of 
the 507 asteroids with taxonomical information, 284 have data in the WISE 
dataset.  Only one object (25285 1998 WB7) has a value of $p_V$ less than 0.1, 
normally associated with C-complex asteroids, and could be considered an albedo 
interloper.  

\begin{figure*}

  \centering
  \begin{minipage}[c]{0.47\textwidth}
    \centering \includegraphics[width=2.8in]{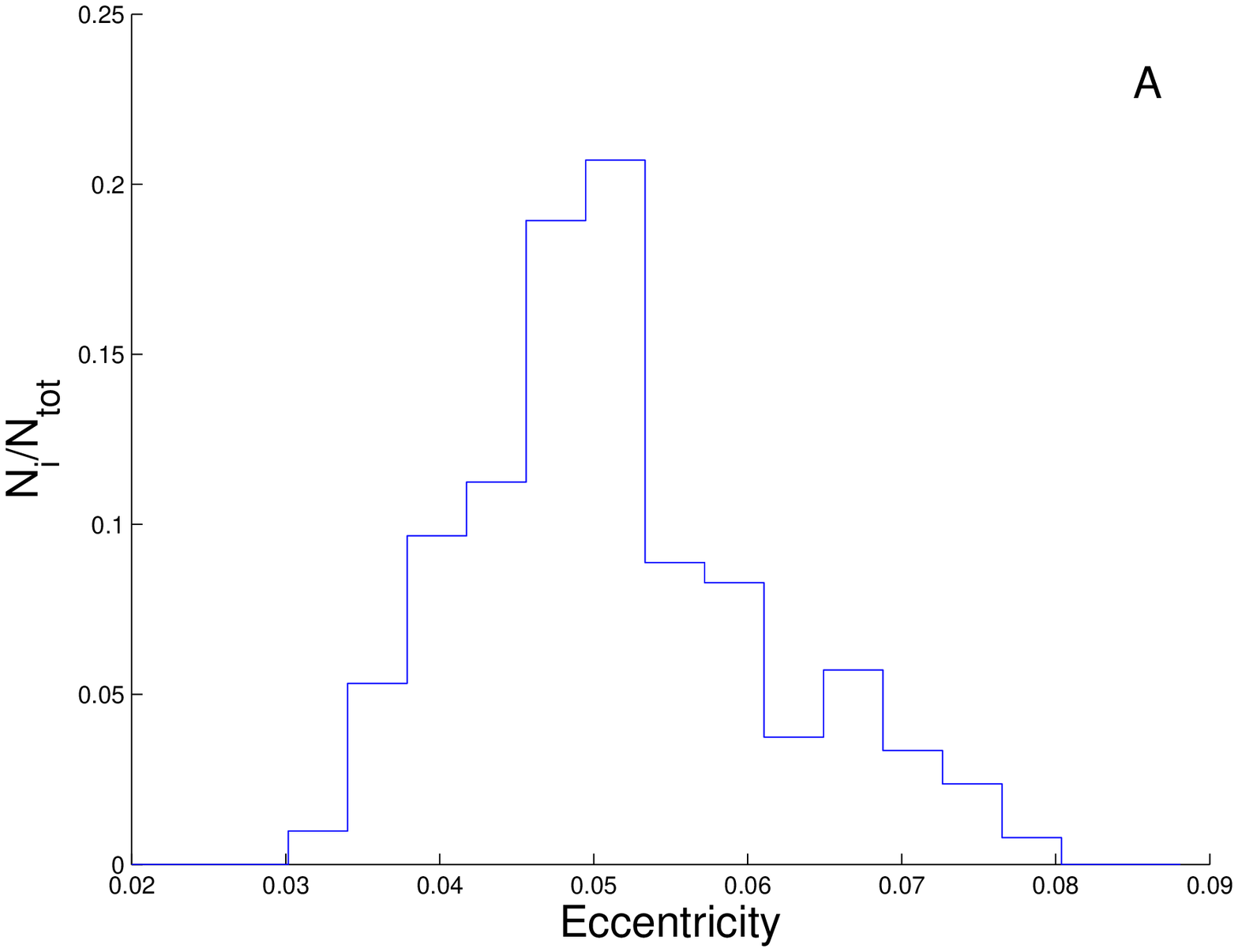}
  \end{minipage}%
  \begin{minipage}[c]{0.47\textwidth}
    \centering \includegraphics[width=2.8in]{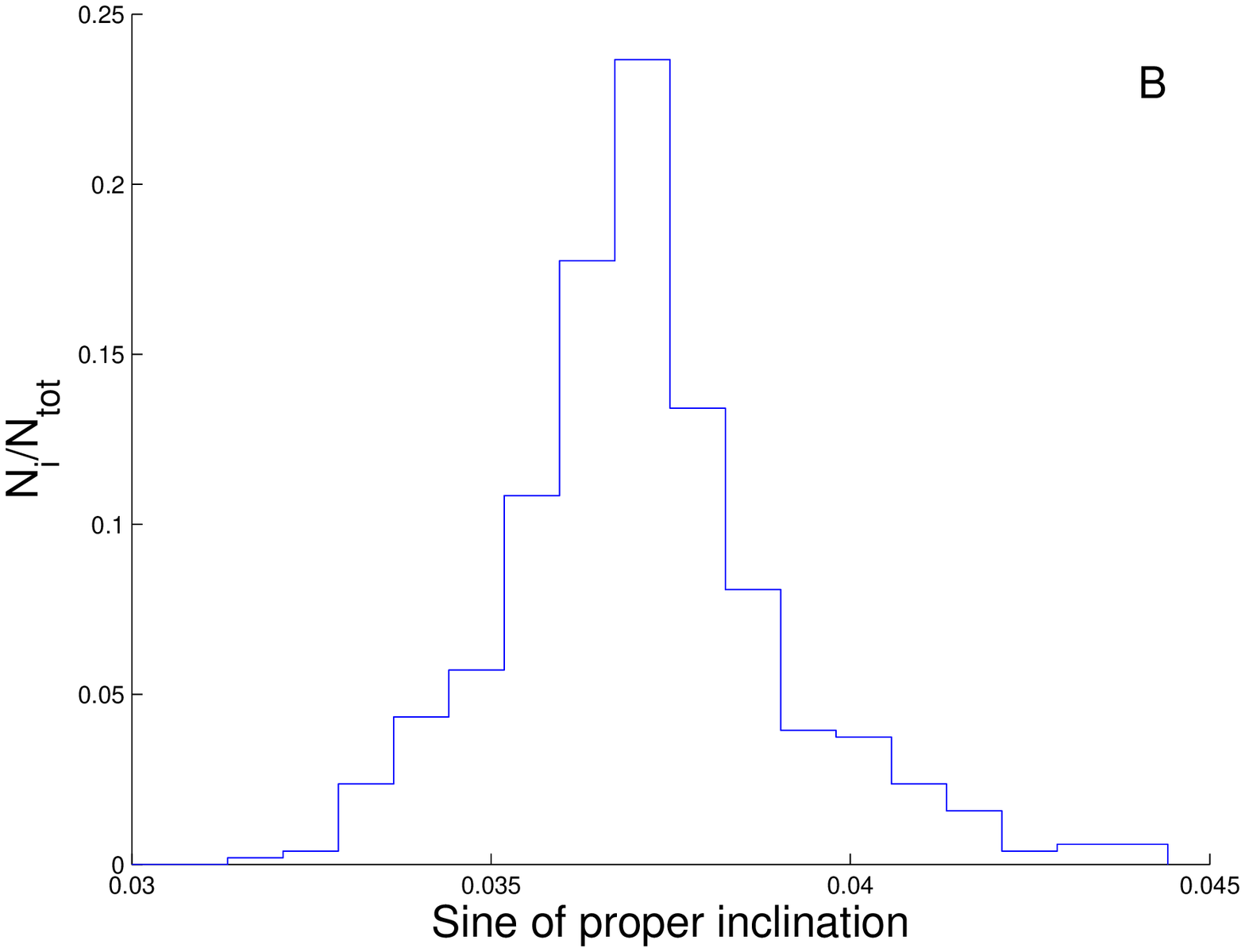}
  \end{minipage}

\caption{Panel A: histograms of the frequency distribution in proper
eccentricity $e$ for a reduced Koronis family with proper semi-major 
axis $a < 2.88$ AU, for objects with $2 < D < 4$ km ($D_3$ population).  
Panel B: similar histograms, but for the distribution in proper 
sine of inclination $\sin(i)$.} 
\label{fig: koronis_e_sini}
\end{figure*}

After eliminating three more objects outside the Yarkovsky isolines of the 
Koronis family (see \citet{Nesvorny_2015}, sect. 4), we were 
left with a sample of 5118 asteroids.  Fig.~\ref{fig: koronis_e_sini}, panel 
A, displays a histogram of the frequency distribution in proper $e$ for a 
reduced Koronis family with proper $a < 2.88$ AU (to avoid including
asteroids that crossed the $g+2g_5-3g_6$ secular resonance and other 
mean-motion resonances, see next section for a more in depth 
discussion of the local dynamics), and, following the approach of 
\citet{Carruba_2015c} objects with $2 < D < 4$~km ($D_3$ population 
hereafter; the sizes of asteroids without WISE data information
where computed using the mean value of the geometric albedo
for the Koronis family from \citet{Nesvorny_2015} and Eq.~1 in 
\citet{Carruba_2003}).  Panel B shows a similar histogram, but 
for the distribution 
in proper $\sin{(i)}$.  Table~\ref{table: e_sin_std} reports the values of 
mean and standard deviation for the different size populations of the 2037 
Koronis sample (about 40\% of the original family) in proper $e$ and 
$\sin{(i)}$.  At smaller sizes the spread in $\sin{(i)}$ increases,
but this essentially stops at sizes lower then 4 km in diameter.

\begin{table}
\begin{center}
\caption{Number of objects, median, and standard deviation for the 
different size populations of the 5118 Koronis sample in proper 
$e$ and $\sin{(i)}$.}
\label{table: e_sin_std}
\resizebox{0.47\textwidth}{!}{
\begin{tabular}{cccc}
\hline
\footnotesize{Population}  & \footnotesize{\# of objects}& \footnotesize{Median $e$} &  \footnotesize{Std(e)} \\ 
\hline
\footnotesize{All}         & \footnotesize{2037}    & \footnotesize{0.0489} 
    & \footnotesize{0.0085}  \\
\footnotesize{$D > 20$~km} &    \footnotesize{10}    & \footnotesize{0.0471} 
    & \footnotesize{0.0030}  \\
\footnotesize{$10 < D < 20$~km} &\footnotesize{44}   & \footnotesize{0.0479} 
    & \footnotesize{0.0039}  \\
\footnotesize{$6 < D < 10$~km}& \footnotesize{126}   & \footnotesize{0.0473}
    & \footnotesize{0.0066}  \\
\footnotesize{$4 < D < 6$~km} & \footnotesize{183}   & \footnotesize{0.0481}
    & \footnotesize{0.0082}  \\
\footnotesize{$2 < D < 4$~km} &\footnotesize{507}   & \footnotesize{0.0495}
    & \footnotesize{0.0097}  \\
\hline
\footnotesize{Population}  & \footnotesize{\# of objects} & \footnotesize{Median $\sin{(i)}$} &  \footnotesize{Std($\sin{(i)}$)} \\ 
\hline
\footnotesize{All}         & \footnotesize{2037}     & \footnotesize{0.0367} 
    & \footnotesize{0.0017}  \\
\footnotesize{$D > 20$~km} &   \footnotesize{10}     & \footnotesize{0.0369}
    & \footnotesize{0.0008}  \\
\footnotesize{$10 < D < 20$~km} &\footnotesize{44}   & \footnotesize{0.0369}
    & \footnotesize{0.0016}  \\
\footnotesize{$6 < D < 10$~km}& \footnotesize{126}   & \footnotesize{0.0371}
    & \footnotesize{0.0014}  \\
\footnotesize{$4 < D < 6$~km} & \footnotesize{183}   & \footnotesize{0.0370}
    & \footnotesize{0.0020}  \\
\footnotesize{$2 < D < 4$~km} & \footnotesize{507}  & \footnotesize{0.0367}
    & \footnotesize{0.0019}  \\
\hline
\end{tabular}}
\end{center}
\end{table}

\begin{figure*}

  \centering
  \begin{minipage}[c]{0.47\textwidth}
    \centering \includegraphics[width=2.8in]{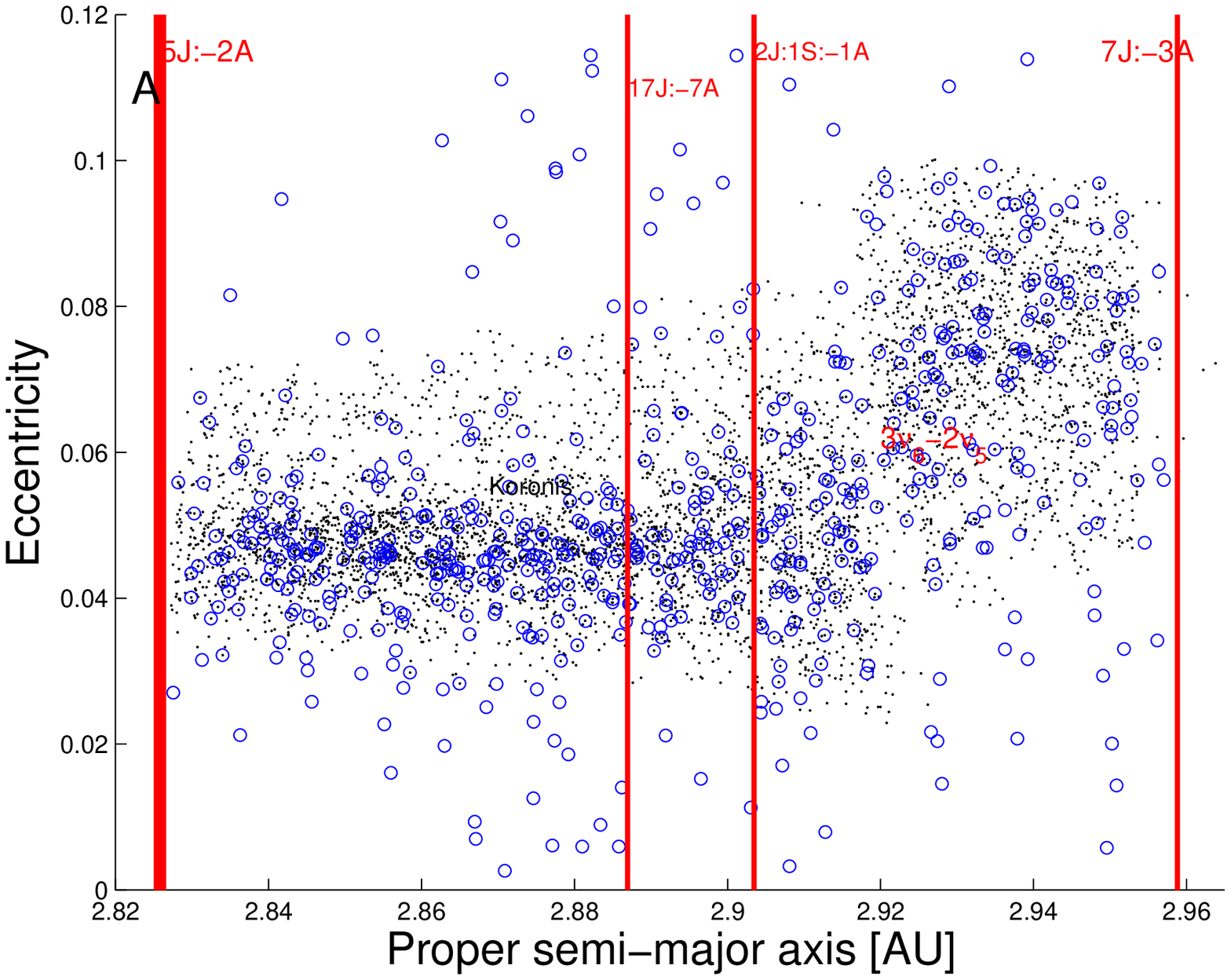}
  \end{minipage}%
  \begin{minipage}[c]{0.47\textwidth}
    \centering \includegraphics[width=2.8in]{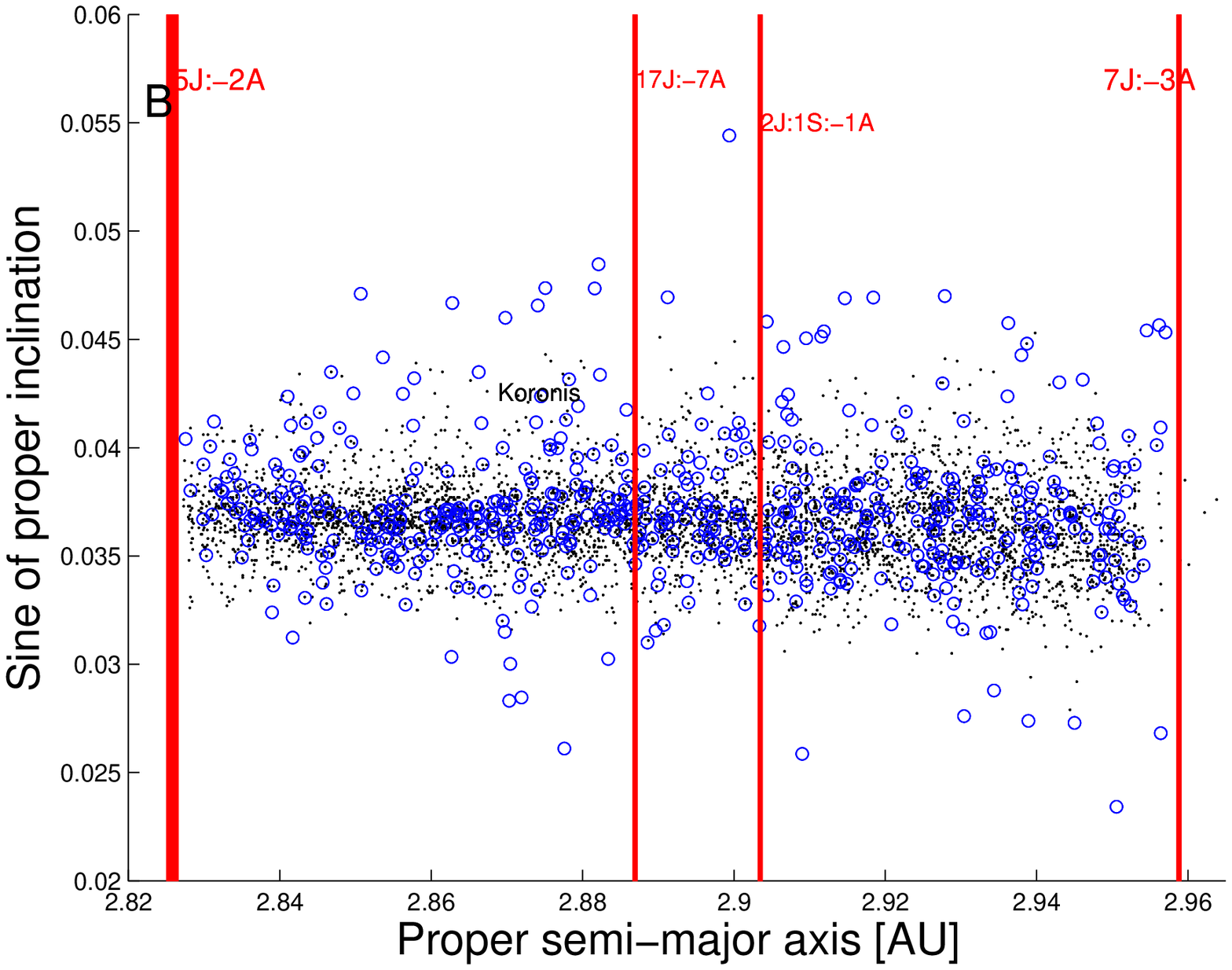}
  \end{minipage}

\caption{An $(a,e)$ (panel A) and $(a,sin(i))$ (panel B) projection
of the HCM (black dots) and SDSS (blue circles) Koronis family members.
Vertical red lines display the location of the main mean-motion resonances
in the region.} 
\label{fig: koronis_halo}
\end{figure*}

\begin{figure*}

  \centering
  \begin{minipage}[c]{0.47\textwidth}
    \centering \includegraphics[width=2.8in]{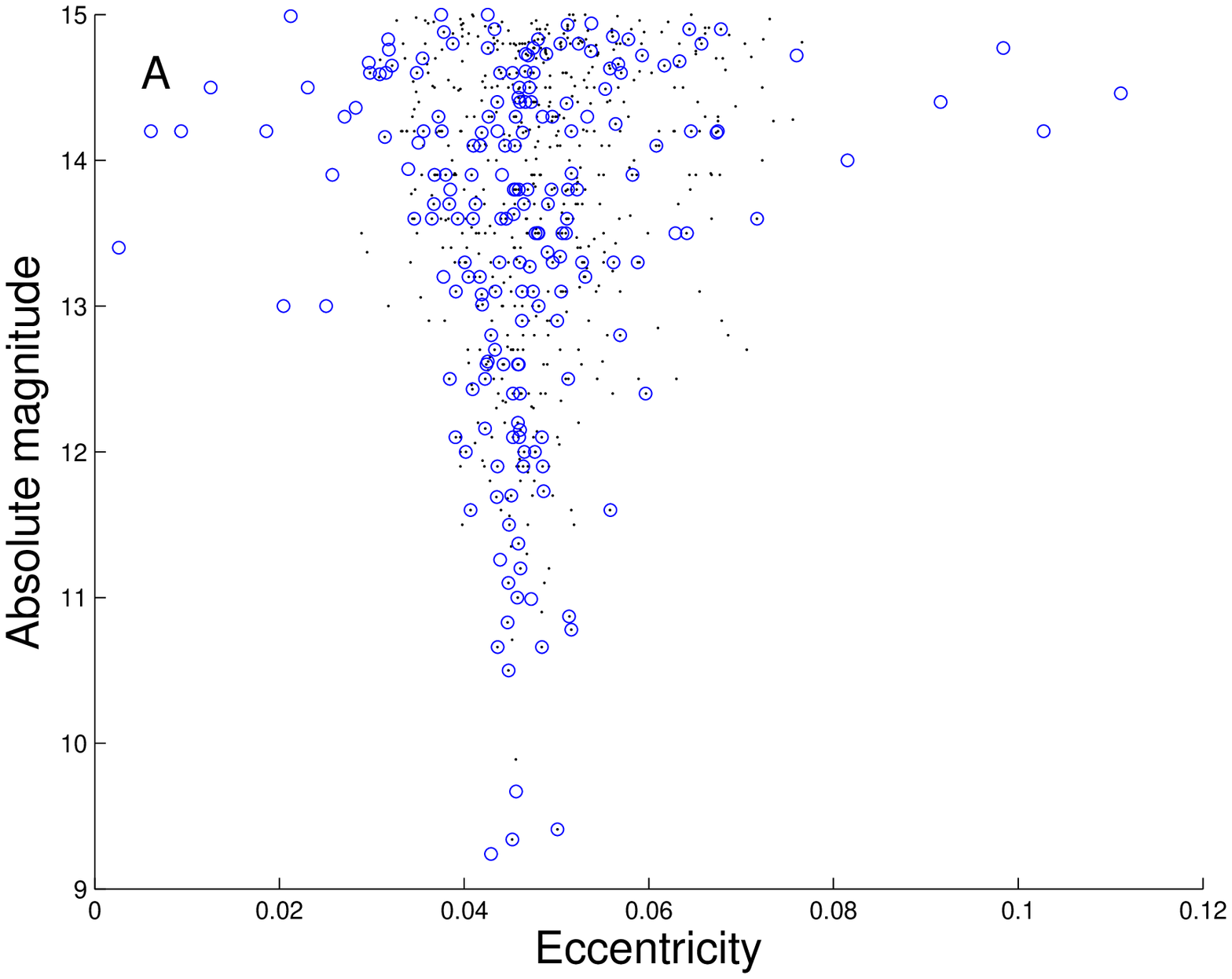}
  \end{minipage}%
  \begin{minipage}[c]{0.47\textwidth}
    \centering \includegraphics[width=2.8in]{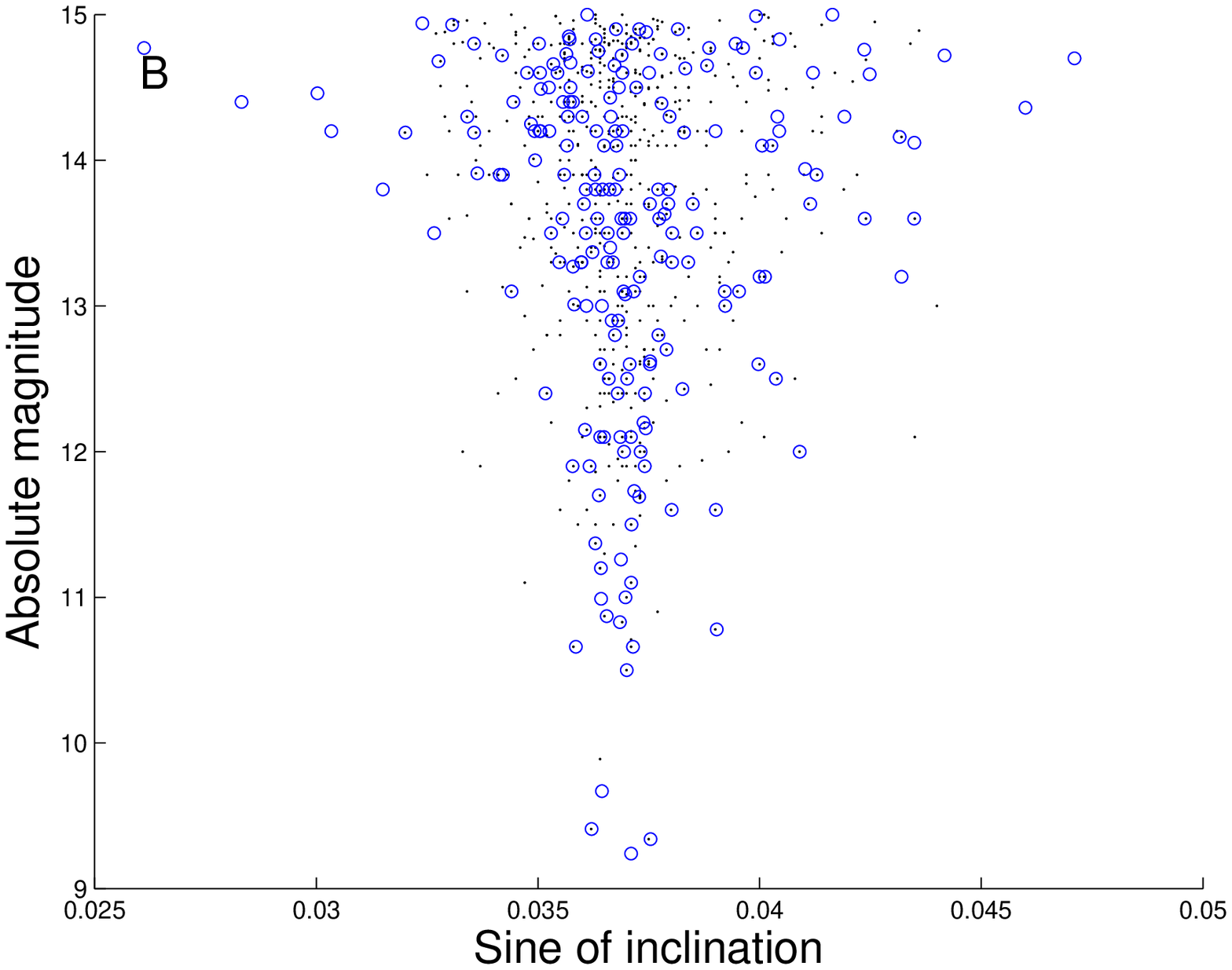}
  \end{minipage}

\caption{An $(e,H)$ (panel A) and $(sin(i),H)$ (panel B) projections
of members of the HCM (black dots) and SDSS (blue circles) Koronis 
family.   The SDSS and, slightly so, the HCM families have 
a characteristic V-shape structure in these domains.} 
\label{fig: koronis_cellino}
\end{figure*}

Rather then being a real characteristics of the Koronis family, we believe
that this is an artifact caused by the method used to identify the 
family in proper element domain.  As discussed in \citet{Broz_Morby_2013} 
and \citet{Carruba_2013}, the hierarchical clustering method (HCM) may 
fail to identify the periphery, or halo, of some large families such as 
Eos and Koronis.  To overcome this limitation, we used the SLOAN and WISE 
data to identify objects in the vicinity of the HCM Koronis family with a 
S-complex taxonomy.  The Koronis family is special among asteroid families 
since it is a S-type family surrounded by a population of C-type objects.
We selected objects whose semi-major axis is between the centers of the 
5J:-2A and 7J:-3A mean-motion resonances, whose $e$ is between 
0 and 0.115 (minimum and maximum values of $e$ of Koronis members $\pm 0.015$, 
\citet{Nesvorny_2015}, and $\sin{(i)}$ is between 0.000 and 0.085 
($\pm 0.04$ from maximum and minimum family range).  After 
eliminating objects with taxonomies in the C-complex, whose value of $p_V$ 
is less than 0.1, and whose values of $e$ and $\sin{(i)}$ differ from
the center of the family by more than 4 standard deviations
of the distribution (see \citet{Carruba_2015c} for more detail on 
the rationale on the use of these criteria, using a $3 \sigma$ criteria
produces changes in the family membership of less than 
1\%, a $5 \sigma$ criteria includes in the Koronis family too many C-type
objects), we obtained a total 
sample of 612 asteroids that could potentially be members of the 
extended SDSS Koronis family, of which 467 are members of the HCM Koronis 
core, as identified in \citet{Nesvorny_2015}, and 145 are in the family
halo. While the number of SDSS objects is inferior to that of the dynamical
HCM Koronis family, and complete only for values of H greater than 13.5 
\citep{DeMeo_2013}, we believe that the possibility of identifying a wider
population in $e$ and $\sin{(i)}$ far outweighs the limitation in 
completeness of the Koronis SDSS sample.

\begin{table}[!htp]
\begin{center}
\caption{Number of objects, median, and standard deviation for the 
different size populations of the 238 Koronis SDSS sample in proper 
$e$ and $\sin{(i)}$.}
\label{table: e_sin_halo}
\resizebox{0.47\textwidth}{!}{
\begin{tabular}{cccc}
\hline
\footnotesize{Population}  &\footnotesize{\# of objects}& 
\footnotesize{Median $e$} &  \footnotesize{Std(e)} \\ 
\hline
\footnotesize{All}         &   \footnotesize{232}    & \footnotesize{0.0469}
     & \footnotesize{0.0148}  \\
\footnotesize{$D > 20$~km} &    \footnotesize{9}    & \footnotesize{0.0468}
     & \footnotesize{0.0032}  \\
\footnotesize{$10 < D < 20$~km}& \footnotesize{19}   & \footnotesize{0.0462}
     & \footnotesize{0.0036}  \\
\footnotesize{$6 < D < 10$~km}& \footnotesize{37}    & \footnotesize{0.0449}   
     & \footnotesize{0.0070}  \\
\footnotesize{$4 < D < 6$~km} &\footnotesize{59}    & \footnotesize{0.0438}
     & \footnotesize{0.0112}  \\
\footnotesize{$2 < D < 4$~km} &\footnotesize{108}    & \footnotesize{0.0439}
     & \footnotesize{0.0190}  \\
\hline
\footnotesize{Population}  &\footnotesize{\# of objects}& 
\footnotesize{Median $\sin{(i)}$} &  \footnotesize{Std(sin(i))} \\ 
\hline
\footnotesize{All}         & \footnotesize{232}      & \footnotesize{0.0370}
& \footnotesize{0.0027}  \\
\footnotesize{$D > 20$~km} &   \footnotesize{9}     & \footnotesize{0.0370}  
& \footnotesize{0.0009}  \\
\footnotesize{$10 < D < 20$~km} &\footnotesize{19}   & \footnotesize{0.0372}  
& \footnotesize{0.0011}  \\
\footnotesize{$6 < D < 10$~km}&  \footnotesize{37}   & \footnotesize{0.0372}
& \footnotesize{0.0012}  \\
\footnotesize{$4 < D < 6$~km} & \footnotesize{59}   & \footnotesize{0.0373}
& \footnotesize{0.0024}  \\
\footnotesize{$2 < D < 4$~km} & \footnotesize{108}   & \footnotesize{0.0368}
& \footnotesize{0.0033}  \\
\hline
\end{tabular}}
\end{center}
\end{table}

\begin{figure}
\centering
\centering \includegraphics [width=0.47\textwidth]{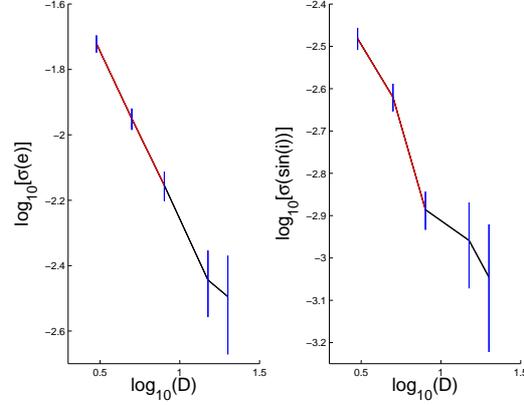}

\caption{Log-log plots of the dependence of the standard deviation
of the distribution in $e$ and $\sin{(i)}$ of the Koronis halo 
members, as a function of diameter.  Vertical blue lines display
the nominal errors, assumed to be inversely proportional to the square 
root of the number of objects in each size bin.  The red line 
connects size bins with with more than 30 objects and nominal errors 
less than 5\%.}
\label{fig: Dsigma_e_sini}
\end{figure}

Fig.~\ref{fig: koronis_halo} displays the $(a,e)$ (panel A) and $(a,sin(i))$ 
(panel B) projections of the HCM (black dots) and SDSS Koronis family
(blue circles). As discussed in \citet{Cellino_2004, Cellino_2009}, and shown 
in Fig.~\ref{fig: koronis_cellino} the SDSS Koronis family displays a V-shape 
distribution in the $(e,H)$ and $(\sin{(i)},H)$ plane, where H is
the asteroid absolute magnitude.  Table~\ref{table: e_sin_halo} displays 
values of mean and standard deviation for the different size populations of the 
halo Koronis sample in proper $e$ and $\sin{(i)}$, that are also shown in 
Fig.~\ref{fig: Dsigma_e_sini}, just for size bins with a population of 
asteroids larger than 30, so as to have a statistical significant sample.
If we assume that errors on the standard deviations in  
proper $e$ and $sin(i)$ are proportional to $\sqrt(n)/n$, with $n$ the 
number of objects in each size bin, then that would correspond to rejecting
samples with a error larger than $\simeq$ 20\%.  By using this approach 
we are consistent with the method of other authors (\citep{Masiero_2011}),
that also did not considered populations of asteroids with $D > 12$ km to avoid
small-number statistics issues.
For the sake of brevity, we will define the $6 < D < 10, 4 < D < 6,$ and
$2 < D < 4$~km size bins as $D_8, D_5$, and $D_3$ populations, respectively. 
We eliminated asteroids with proper $a > 2.88$ AU to 
avoid including objects that crossed the $g+2g_5-3g_6$ secular resonance
and other local mean-motion resonances.  To estimate the contribution
of background objects to the SDSS Koronis family, following the approach
of \citet{Broz_Morby_2013} we counted the number of S-complex objects
in boxes in the $(a,e)$ and $(a,\sin{(i)})$ planes, limited in $a$ by the 5J:-2A
mean-motion resonance and $a < 2.88$ AU, and in $e$ and $\sin{(i)}$ by the 
center of the distribution plus 4 $\sigma$ values of the real family as 
the minimum value, and this quantity plus the length of the distribution 
in $e$ and $\sin{(i)}$ of the real family as the maximum value. The maximum 
number of S-complex objects that we found in these boxes was 6 objects, 
which corresponds to 2.1\% of the population of the Koronis SDSS family in 
that area (238 asteroids).   Overall, this seems to suggest that our 
determination of the Koronis family halo should not be too much affected by 
background contributions.  Also, possible errors caused by the (maximum) 
5\% photometric noise in the SDSS-MOC4 data used in the \citet{DeMeo_2013} 
taxonomic identification method should affect at most 1\% of the SDSS 
Koronis family S-complex objects, according to our estimates.

Distributions in $e$ and $\sin{(i)}$ became more spread at smaller
sizes, and in a steeper way with respect to what found for the 
nominal HCM core of the Koronis family.  The dependence of $\sigma{(e)}$ and 
$\sigma{(sin(i))}$ with respect to $D$ is similar.  Assuming that 
$\sigma$ follows a power-law of the form $\sigma = C (\frac{1}{D})^{\alpha}$, 
best-fit values of $\alpha$ for size bins with more than 30 objects, 
red line in Fig.~\ref{fig: Dsigma_e_sini}, are 1.02 and 0.95 for the $e$ 
and $\sin{(i)}$ 
distributions, with characteristic errors of approximately 9\% and 3\%, 
respectively.  Both distributions seem to be inversely proportional with 
respect to asteroid diameters (i.e., $\sigma = C \frac{1}{D}$).  
\section{Local dynamics}
\label{sec: loc_dyn}
To better understand the importance of the local web of mean-motion 
and secular resonances, we obtained dynamical maps in the domain of
synthetic proper $(a,e)$ and $(a,sin(i))$ with the method of 
\citet{Carruba_2010}.  Dynamical maps do not account for 
non-conservative forces such as the Yarkovsky force, but are 
useful to identify the location of the main mean-motion and secular 
resonances in proper element domains.
We used 1775 particles in the osculating $(a,e)$ plane,
using a step of 0.002~AU in $a$ and 0.005 in $e$, with 71 intervals
in $a$ and 25 in $e$ starting at $a = 2.82$~AU and $e=0$, respectively.
In the osculating $(a,sin(i)$ plane we used the same intervals in $a$,
and 51 intervals of 0.06$^{\circ}$ in osculating $i$, starting at 
$i = 0^{\circ}$, for a total of 3621 particles.  Test particles
were integrated over 20 Myr under the influence of all planets,
except Mercury (whose presence was accounted for as a barycentric
correction to the mass of the Sun), with a time step of 20 days.

\begin{figure*}[!htp]

  \centering
  \begin{minipage}[c]{0.47\textwidth}
    \centering \includegraphics[width=2.8in]{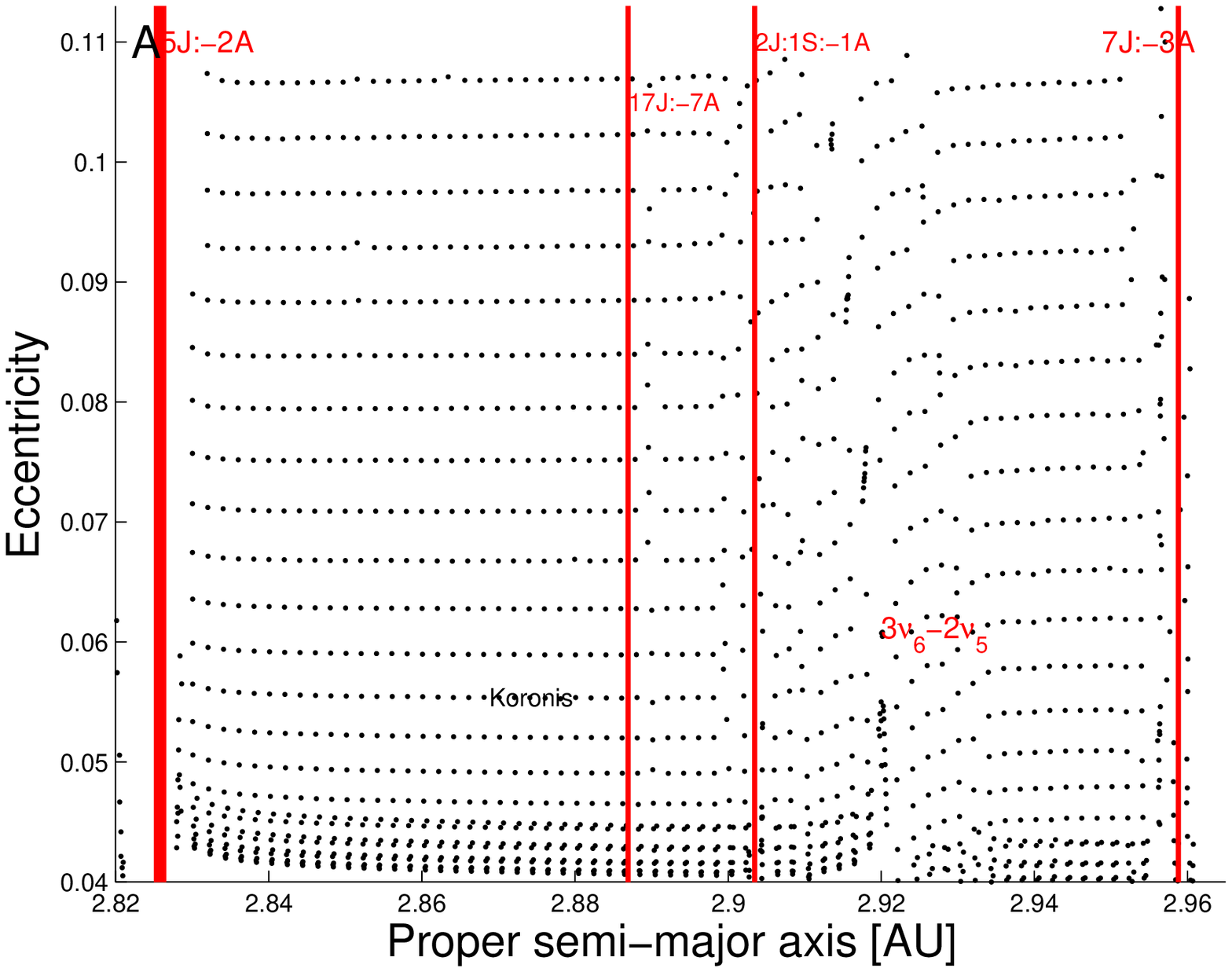}
  \end{minipage}%
  \begin{minipage}[c]{0.47\textwidth}
    \centering \includegraphics[width=2.8in]{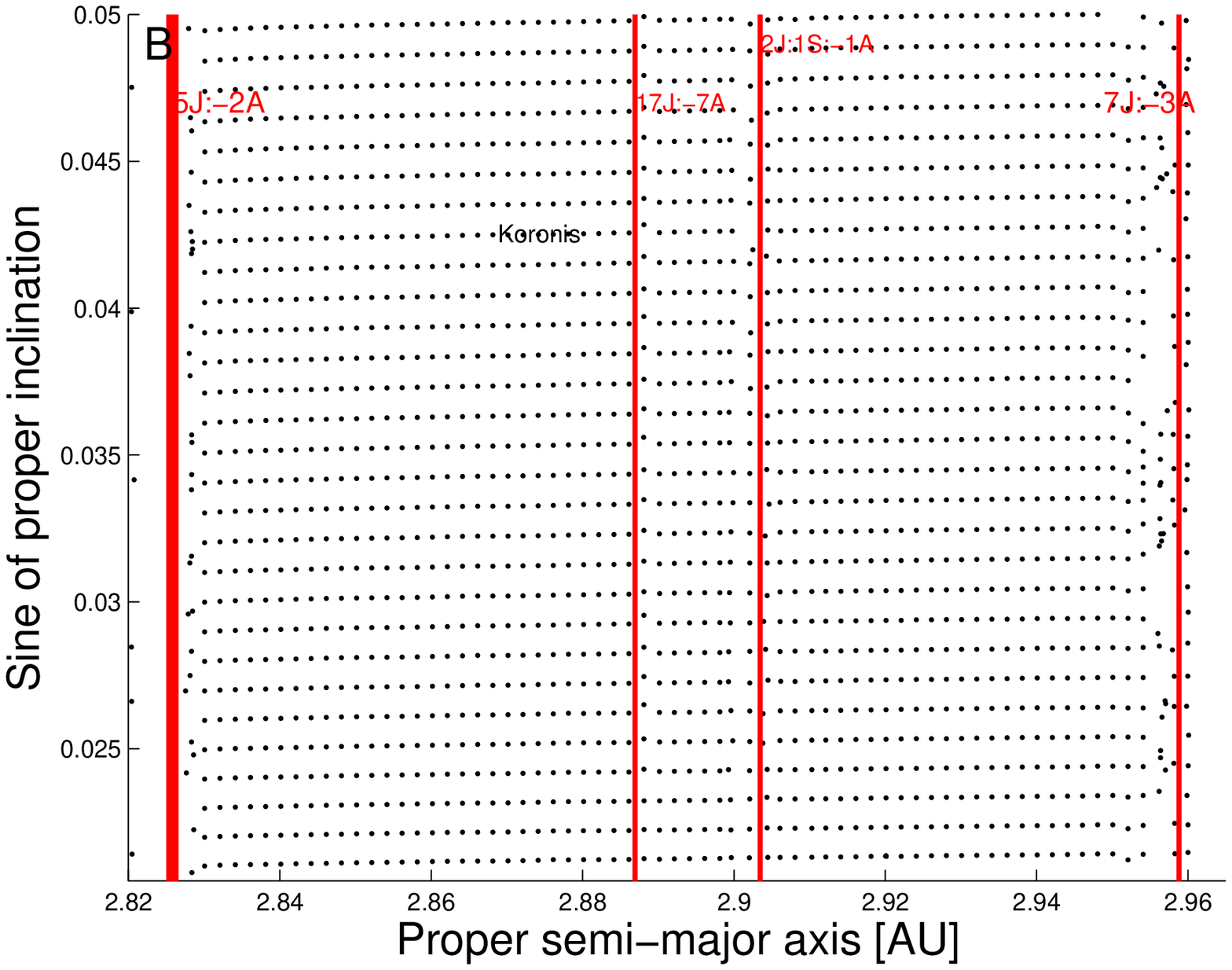}
  \end{minipage}

\caption{A dynamical map in the domain of proper $(a,e)$ (panel A) and 
proper $(a,sin(i))$ (panel B) for the region of the Koronis family. 
Black dots display the location in the two proper elements domains of
each test particle that survived for the length of the integration.
Vertical red lines display the location of the local mean-motion resonances.} 
\label{fig: Koronis_dyn}
\end{figure*}

Fig.~\ref{fig: Koronis_dyn} shows our results in the proper $(a,e)$ 
(panel A) and proper $(a,sin(i))$ (panel B) plane.  Mean-motion resonances
appear as vertical strips devoid of particles, secular resonances cause
particles to be aligned in inclined bands.  Apart from the well-know
5J:-2A and 7J:-3A mean-motion resonances, we also identified the 17:-7A
two-body resonance and the 2J:1S:-1A three-body resonance in the region
(a minor role is also played by the 6S:-1A two body resonance with Saturn,
near the 2J:1S:-1A three-body resonance).   No major secular resonances appear 
in the $(a,sin(i))$ plane. In the $(a,e)$ plane, however, one can notice 
the strong effect of the $3{\nu}_6-2{\nu}_5$ secular resonance, that appears 
as an inclined band at $\simeq 2.92$~AU.

To check if all important secular resonances were identified in the dynamical 
maps, we also computed the orbital location of all secular resonances whose 
combination of proper $g$ and $s$ is within the values covered by the 
Koronis family, and checked the number of likely resonators
\citep{Carruba_2009} for each resonance (likely resonators are defined
as the objects whose combination of asteroidal proper frequencies is within
$\pm 0.3$ arcsec/yr from the resonance center. For the case of the 
$z_1 = g-g_6 + s-s_6$ resonance, this would correspond to $g+s = g_6+s_6 = 1.898$
arcsec/yr; the actual threshold may vary for higher order resonances, but 
the 0.3 arcsec/yr boundary usually provide a good first order of magnitude
criteria).  Not all likely resonators are in librating states, 
but the number of these objects may provide a first clue on the 
dynamical strength of each resonance.

\begin{table}[!htp]
\begin{center}
\caption{Main secular resonances in the Koronis region, frequency
    value, and number of likely and actually resonant asteroids.}
\label{table: sec_res}
\resizebox{0.48\textwidth}{!}{
\begin{tabular}{cccc}
\hline
\footnotesize{Resonance} & \footnotesize{Frequency value} & 
\footnotesize{Likely}      \\
\footnotesize{argument} &  \footnotesize{$[``/yr]$} & 
\footnotesize{resonators}  \\  
\hline
                 &  \footnotesize{g resonances}     &             \\
\footnotesize{$g-3g_6+2g_5$}&  \footnotesize{76.215}  &  \footnotesize{167} \\  
                 &  \footnotesize{s resonances}     &             \\ 
\footnotesize{$s-s_6+2g_6-2g_5$} & \footnotesize{-74.317} & \footnotesize{42} \\
\hline
\end{tabular}}
\end{center}
\end{table}

Table~\ref{table: sec_res} shows the results for the two
secular resonances with a number of likely resonators larger
than 1: the $3{\nu}_6-2{\nu}_5 = g-3g_6+2g_5$ resonance, already described 
in the seminal paper of \citet{Bottke_2001}, and the 
$2{\nu}_5-2{\nu}_6+{\nu}_{16} = s-s_6+2g_6-2g_5$ $s-$type
resonance, that is near the 7J:-3A mean-motion resonance and has therefore
a limited importance in affecting the dynamical evolution of the Koronis 
group.  The very limited number of secular resonances with a significant
population of likely resonators found in this region
confirms our initial hypothesis that the Koronis family lies in a relatively
dynamically quiet region.

How much the local dynamics can be responsible for the current dispersion
in proper $e$ and $\sin{(i)}$ of the Koronis family?   To answer this question,
we performed simulations with the $SYSYCE$ integrator 
(Swift$+$Yarkovsky$+$Stochastic YORP$+$Close encounters) of 
\citet{Carruba_2015a}, modified to also account for past changes in 
the values of the solar luminosity.   The numerical set-up of 
our simulations was similar to 
what was discussed in \citet{Carruba_2015a}: we used the optimal values of the 
Yarkovsky parameters discussed in \citet{Broz_2013} for S-type asteroids
(the spectral type of most Koronis family members),
the initial spin obliquity was random, and normal reorientation timescales 
due to possible collisions as described in \citet{Broz_1999} were 
considered for all runs.  We warn the reader that using other values of 
key parameters, such as the bulk density and thermal conductivity of 
asteroids may significantly alter the strength of the Yarkovsky force
\citep{Masiero_2012}.  For a review of the effect of changing these parameters
on the estimated age of the Koronis family please see \citet{Carruba_2015b}. 
We integrated our test particles  
under the influence of all planets, and obtained synthetic proper elements
with the approach described in \citet{Carruba_2010}.  Initial conditions
for the 502 test particles for the Koronis family were obtained with the 
approach described in \citet{Carruba_2015b}, i.e, we generated 
a fictitious family with the ejection parameter $V_{EJ}$ equal to that
obtained from our Monte Carlo simulations of the Koronis family, i.e.,
60 m/s.  The size distribution of the test particles used followed a 
size-frequency distributions (SFD) with 
an exponent $-\alpha$ that best-fitted the cumulative distribution equal 
to 3.6, a fairly typical value \citep{Masiero_2012}, and with diameters in 
the range from 2.0 to 12.0 km.

\begin{figure*}

  \centering
  \begin{minipage}[c]{0.47\textwidth}
    \centering \includegraphics[width=2.8in]{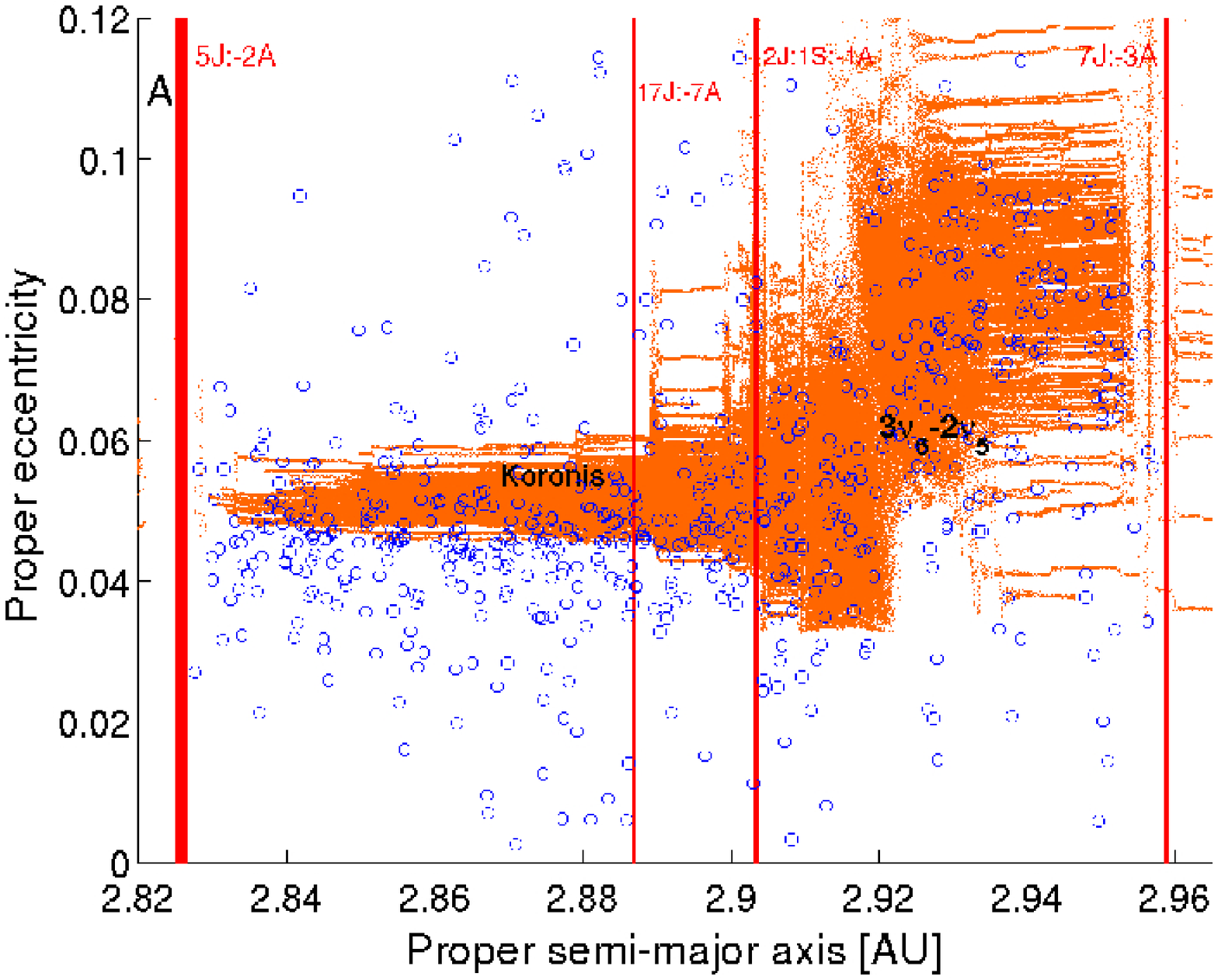}
  \end{minipage}%
  \begin{minipage}[c]{0.47\textwidth}
    \centering \includegraphics[width=2.8in]{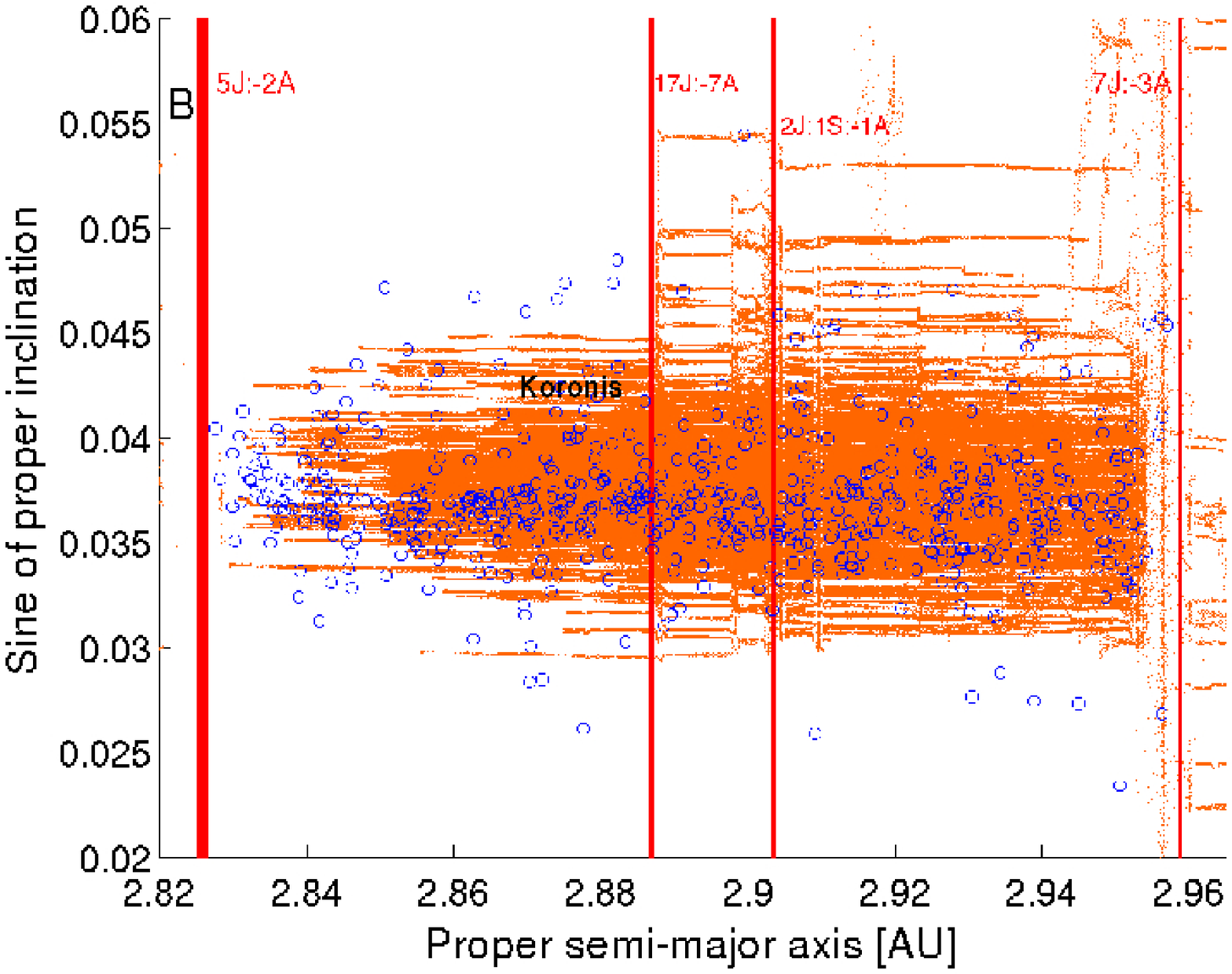}
  \end{minipage}

\caption{Dynamical evolution of test particles (dark orange dots) in the proper 
$(a,e)$ (panel A) and proper $(a,sin(i))$ (panel B) planes for the Koronis 
family. Blue circles identify the orbital location of SDSS members, 
the other symbols are the same as in Fig.~\ref{fig: koronis_halo}.}
\label{fig: Koronis_yarko}
\end{figure*}

Fig.~\ref{fig: Koronis_yarko} displays the dynamical evolution of our test
particles in the proper $(a,e)$ (panel A) and proper $(a,sin(i))$ (panel B) 
planes (each orange path represent values of proper element for 
each given particle).  Blue circles identify the orbital location of 
SDSS Koronis members, vertical red lines display the location of
mean-motion resonances. As in \citet{Bottke_2001} we observe the very 
relevant effect of the $3{\nu}_6-2{\nu}_5$ secular resonance in increasing 
values of proper eccentricities.  We were not able, however, to
produce the low-eccentricity population of asteroids at $a < 2.88$ au.
This may be caused by either i) the fact that the family identified in 
SDSS data may be too extended
in eccentricity, ii) some other mechanism of dynamical mobility in proper
$e$ not accounted in our model, such as close encounters with massive 
asteroids and dwarf planets \citep{Carruba_2003} or pericenter 
secular resonances with Ceres \citep{Novakovic_2015} could have been at play, 
or iii) that our simulations did not account in a large
enough manner for reorientations events of particles across mean-motion 
and secular resonances that may have caused further spread in eccentricity
(see also the discussion in the next section).  Since the goal of this paper 
was to concentrate on the inclination distribution of the Koronis family, 
we believe this is an acceptable trade-off.  But explaining the current
distribution in proper $e$ of the Koronis family certainly remain a 
challenge for future research.  A few particles interacted with the
$2{\nu}_5-2{\nu}_6+{\nu}_{16}$ nodal resonance near the 7J:-3A mean-motion
resonance and were scattered to higher values of inclination, as a result.

To compute how much local dynamics influenced the dispersion in proper 
$e$ and $\sin{(i)}$, we calculated the time behavior of the standard deviation 
of $\delta (e)$ and $\delta [\sin{(i)}]$ for all 502 particles.  Standard 
deviations of changes in $e$ and $\sin{(i)}$ with respect to their initial 
values were computed so as to eliminate the effect of the assumed 
initial dispersion and obtain an estimate of changes caused by dynamics.  We 
did not consider in our computation particles that escaped from the region 
of the Koronis family, defined as a box given by the current maximum and 
minimum values in $(a,e,\sin{(i)})$.

\begin{figure*}[!htp]

  \centering
  \begin{minipage}[c]{0.47\textwidth}
    \centering \includegraphics[width=2.8in]{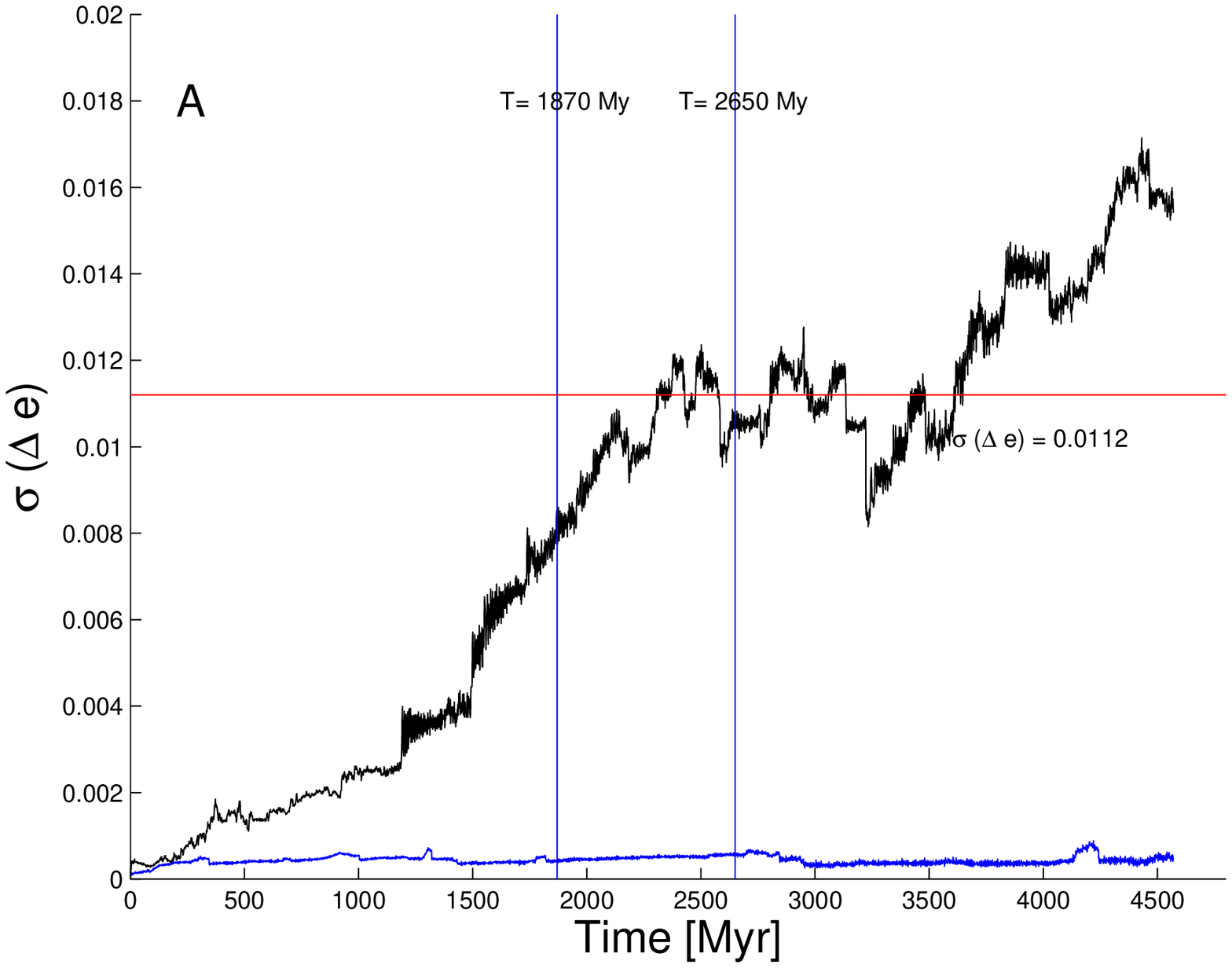}
  \end{minipage}%
  \begin{minipage}[c]{0.47\textwidth}
    \centering \includegraphics[width=2.8in]{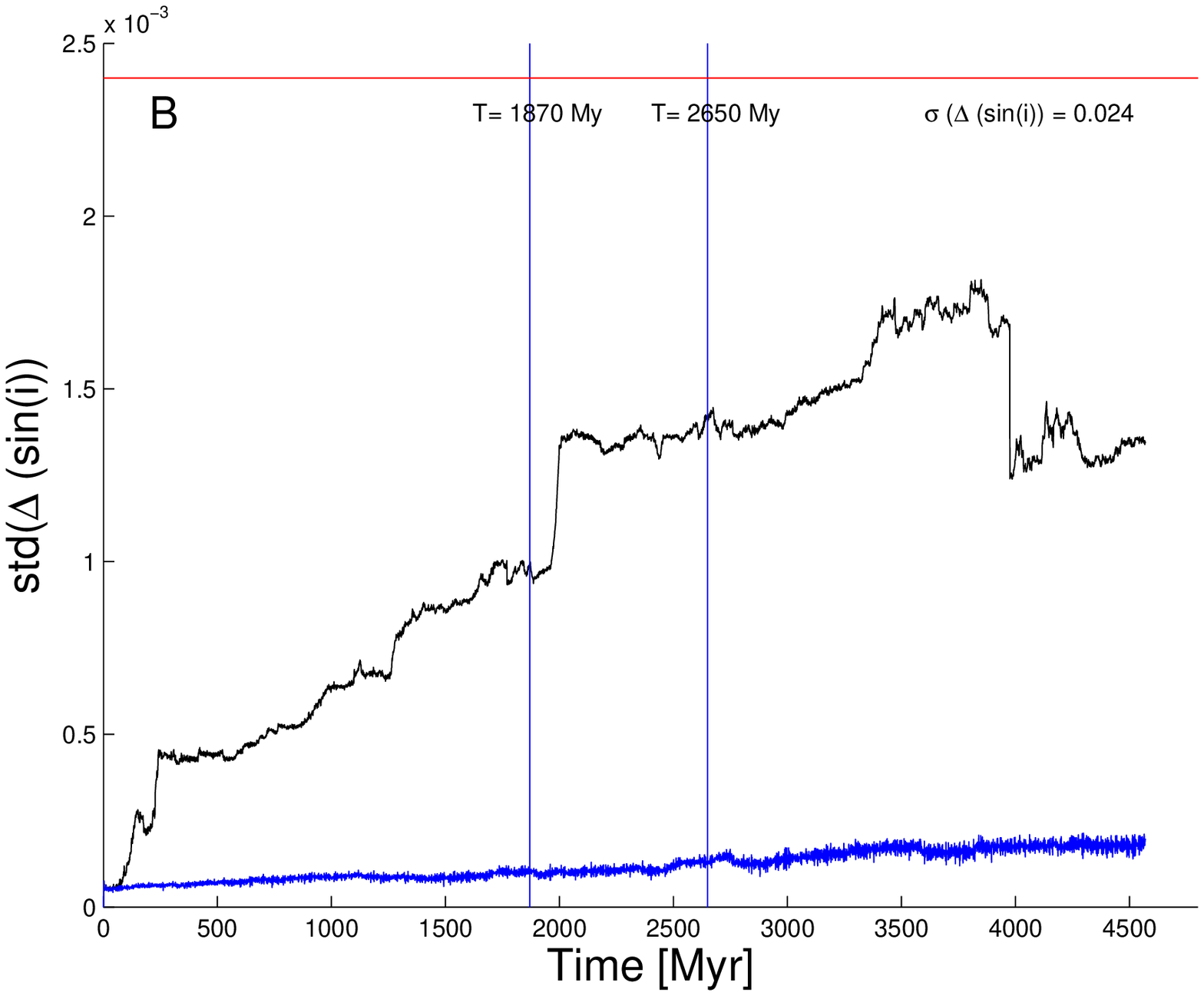}
  \end{minipage}

\caption{Temporal evolution of the standard deviations in proper 
$\delta (e)$ (panel A) and $\delta [\sin{(i)}]$ (panel B) for 
the particles in the size range from 4 to 6 km ($D_5$) in our simulation. 
The black line refers to the whole Koronis sample, while the blue line
is just for the $a < 2.88$~au population.  Vertical lines  display 
the minimum  and maximum ages of the family, as estimated in 
\citet{Carruba_2015b} using a Yarko-YORP method to fit the distribution in 
proper semi-major axis.   The horizontal red line displays the current value
of dispersion in proper $e$ and $\sin{(i)}$ for the $a < 2.88$~au population, 
as from Table.~\ref{table: e_sin_halo}.}
\label{fig: Koronis_dedi}
\end{figure*}

Fig.~\ref{fig: Koronis_dedi} displays our results, for particles in the
size range from 4 to 6 km, while Table~\ref{table: dedi_std_fict} summarizes 
different values for three size interval distributions ($D_8, D_5$, and 
$D_3$ populations).  Changes in $\delta (e)$ and $\delta [\sin{(i)}]$ in 
the region with $a < 2.88$~au are small, and amount at most to just 5\% of 
the currently observed values of standard deviations in $\delta (e)$ 
and $\delta [\sin{(i)}]$.  If we consider the whole Koronis family, however, 
changes in $\delta (e)$ are of the order of 93\%, and in $\delta [\sin{(i)}]$ 
of 50\% for the Koronis SDSS sample with $4 < D < 6$ km, respectively.  
Overall, these results suggest that a significant part of the spread in 
proper $\sin{(i)}$ (and less so for proper $e$) for the $a < 2.88$ au Koronis 
population could still bear traces of the original ejection speeds
\footnote{To check how robust and model-independent are our results, we 
compared the time evolutions the standard deviation of changes in proper 
$e$ and $\sin(i)$ obtained by our integration with those obtained by 
\citet{Bottke_2001} in their seminal work on the Yarkovsky effect and 
the Koronis family.  In their model there were no reorientations, all 
particles had 4 km diameters, spin axes were randomly distributed and fixed, 
and the integration lasted $\simeq$600 Myr, under the influence of all 
planets.  Despite the differences in which non-gravitational forces were 
modeled, we found that i) changes in standard deviation of $\delta e$ and 
$\delta [\sin(i)]$ were essentially negligible in the $a < 2.88$ au region of 
the Koronis family, as observed in our simulations (blue lines in 
Fig.~\ref{fig: Koronis_dedi}), and ii) differences for the whole Koronis 
family were at most of the order of 20\%.  
Since standard deviations in $\delta e$ and $\delta [\sin(i)]$ 
produced by dynamics sum quadratically with those from the current
observed distribution (see Eq.~\ref{eq: sigma_in}), a 
20\% difference corresponds to a 4\% error, which is acceptable, 
in our opinion.}.  
\begin{table}
\begin{center}
\caption{Standard deviations of changes in proper $e$ and $\sin{(i)}$
at the minimum and maximum age estimate for the Koronis family, for
three size distributions of the simulated dynamical group for the whole
family and for $a < 2.88$~au population.}
\label{table: dedi_std_fict}
\resizebox{0.47\textwidth}{!}{
\vspace{0.5cm}
\begin{tabular}{ccccc}
\hline
\footnotesize{Size int.}&  \footnotesize{Min} & \footnotesize{Max}  & 
\footnotesize{Min} &  \footnotesize{Max}  \\
\footnotesize{$[km]$} &  {\small $\sigma [\delta (e)]$} & 
\footnotesize{$\sigma [\delta (e)]$} & 
\footnotesize{$\sigma [ \delta (\sin{(i)})]$} & 
\footnotesize{$\sigma [\delta (\sin{(i)})]$} \\
\hline
\footnotesize{Whole sample} & & & & \\
\hline
$D_8$ & 0.0051 & 0.0063 & 0.0006 & 0.0009 \\
$D_5$ & 0.0085 & 0.0104 & 0.0010 & 0.0014 \\
$D_3$ & 0.0099 & 0.0169 & 0.0015 & 0.0021 \\
\hline
\footnotesize{$a < 2.88~au$ pop.} & & & & \\
\hline
$D_8$ & 0.0002 & 0.0003 & 0.0001 & 0.0001 \\
$D_5$ & 0.0004 & 0.0006 & 0.0001 & 0.0001 \\
$D_3$ & 0.0006 & 0.0008 & 0.0001 & 0.0002 \\
\hline
\end{tabular}}
\end{center}
\end{table}

This could also explain the dependence of standard 
deviations in $e$ and $\sin{(i)}$ as a function of the asteroid diameter, found 
in Sect.~\ref{sec: fam_ide}.  Two values of speed can be computed for the 
fragments of a collision.  One defines the ejection speeds of the objects 
immediately after the collision.  The second one characterizes the speed
after the fragments escape the gravitational pull of the parent body,
or velocity at infinity.  In this work we will define the first
speed as initial ejection speed, and the latter as terminal ejection speed,
so as to avoid the cumbersome expression ``ejection speed at infinity''.
If one assumes that the initial ejection velocity field follows a Gaussian 
distribution of zero mean and standard deviation given by 
\citep{Vokrouhlicky_2006}:

\begin{equation}
{\sigma}_{V_{ej}}=V_{EJ}\cdot \frac{5km}{D},
\label{eq: std_veJ}
\end{equation}

\noindent 
where $D$ is the body diameter in km, and $V_{EJ}$ is a parameter 
describing the width of the initial velocity field, then the dependence of 
${\sigma}_{V_{ej}}$ on $D$ should be inverse.  \citet{Carruba_2015c}
described how ejection speeds are related to distributions in proper
$e$ and $\sin{(i)}$.  If, as suggested by the results of our simulations,
the current distribution in $e$ and $i$ of the Koronis family should have been
less affected by dynamics than those of proper $a$, at least for
$a < 2.88$ au, then one would expect that, to some limits, the 
dependence of ${\sigma}_{e}$ and ${\sigma}_{sin(i)}$ on $D$ should still be 
roughly inverse, as observed.  We will further investigate this issue in the
next section.
\section{Ejection velocity field}
\label{sec: ej_field}

Proper orbital elements can be related to the components of 
terminal ejection velocity along the direction of 
orbital motion ($\delta v_t$), in the radial direction ($\delta v_r$), and 
perpendicular to the orbital plane ($\delta v_W$) through the Gauss equations 
\citep{Murray_1999}:

\begin{small}
\begin{equation}
\frac{\delta a}{a} = \frac{2}{na(1-e^2)^{1/2}}[(1+ecos(f) \delta v_t +
(e sin(f)) \delta v_{r}],
\label{eq: gauss_1}
\end{equation}

\begin{equation}
\delta e =\frac{(1-e^2)^{1/2}}{na}\left[\frac{e+cos(f)+e cos^2 (f)}{1+e cos (f)}
\delta v_t+sin(f) \delta v_r\right],
\label{eq: gauss_2}
\end{equation}

\begin{equation}
\delta i = \frac{(1-e^2)^{1/2}}{na} \frac{cos(\omega+f)}{1+e cos(f)} \delta v_W. 
\label{eq: gauss_3}
\end{equation}
\end{small}
\noindent 
where $\delta a = a-a_{ref}, \delta e = e-e_{ref}, \delta i= i-i_{ref}$, 
where $a_{ref}, e_{ref}, i_{ref}$ define a reference orbit and $f$ and 
$\omega$ are the true anomaly and perihelion argument of the disrupted 
body at the time of impact.  Since proper $a$ is affected by 
non-gravitational forces such as the Yarkovsky and YORP effects, 
it is not therefore possible to use the first equation to obtain 
information on the primordial values of the components of the 
terminal velocities.  In this section we focus our attention on the proper 
$e$ and $i$ distribution, for $a< 2.88$ au.

Concerning values of $e_{ref}$ and $i_{ref}$, since the Koronis family 
originated from a catastrophic disruption
event that left no main largest fragment \citep{Nesvorny_2015}, 
we obtained an estimate of the barycenter position in the
$(a,e,sin(i))$ domain.  Apart for the few asteroids for which a mass
determination was available in \citet{Carry_2012}, we estimated the masses
of the other objects using the WISE values of diameters and the 
density of 243 Ida, the only member of Koronis visited by the 
space mission Galileo, as reported in \citet{Carry_2012}.  Values
of $a_b,e_b,\sin{(i_b)}$ at the barycenter were then obtained by means of
a weighted average, with the weigth on the asteroid proper
elements given by each asteroid estimated mass, divided by the total 
mass of the family.
\begin{figure*}
\centering
  \begin{minipage}[c]{0.45\textwidth}
    \centering \includegraphics[width=2.6in]{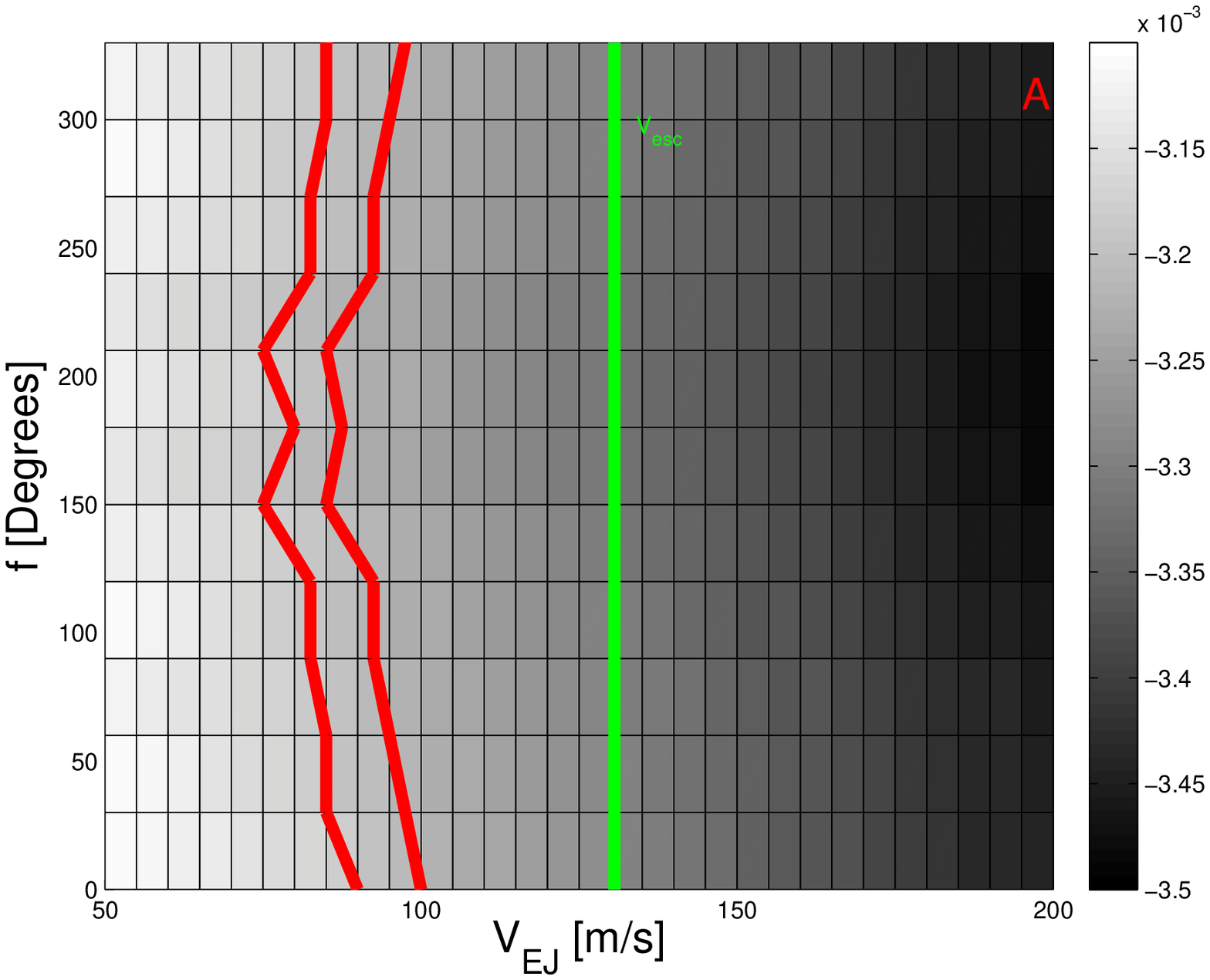}
  \end{minipage}%
  \begin{minipage}[c]{0.45\textwidth}
    \centering \includegraphics[width=2.6in]{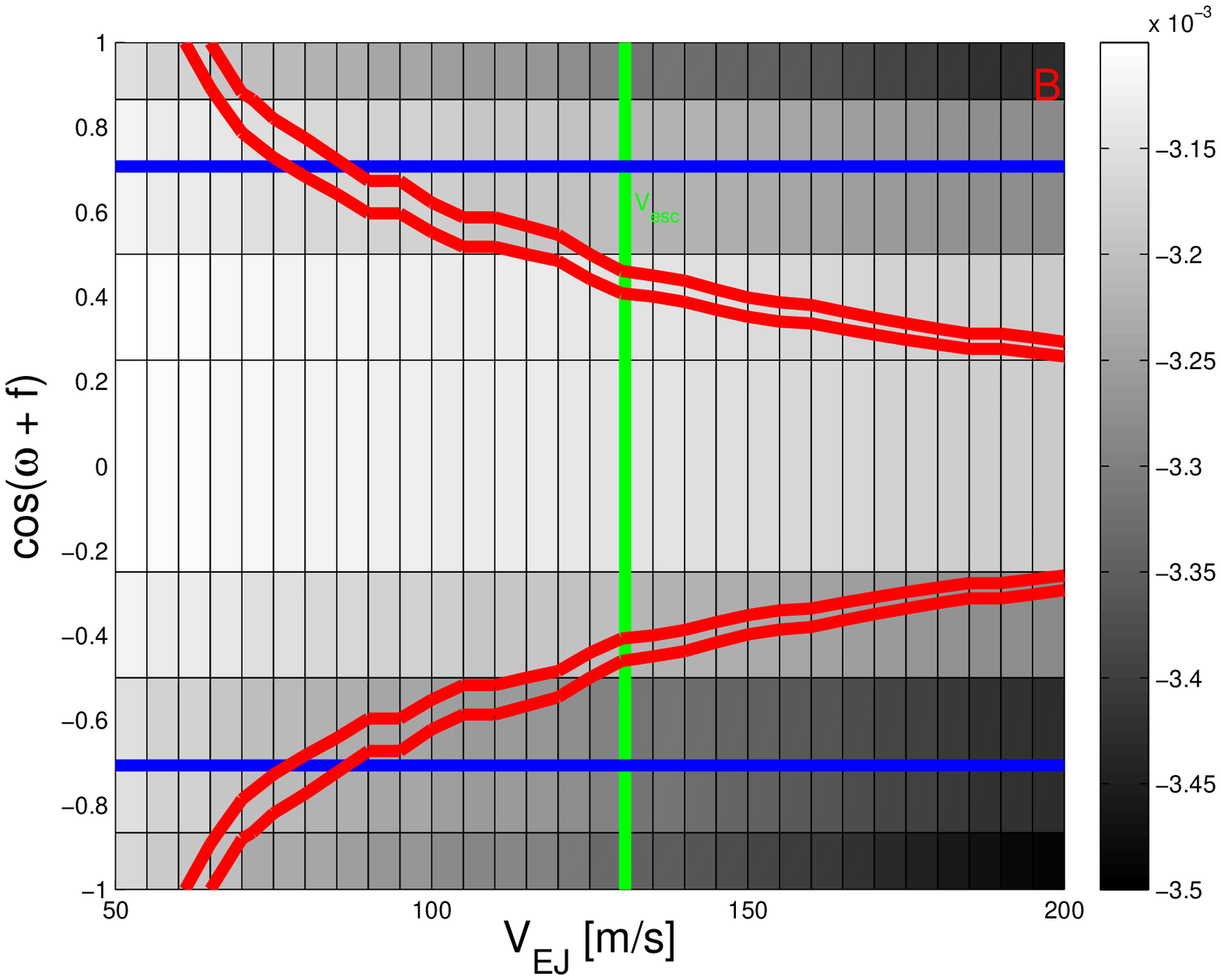}
  \end{minipage}

\caption{Values of ${\sigma}_{sin(i)}$ as a function of $f$ and $V_{EJ}$
(panel A) and $\cos{(\omega +f)}$ and $V_{EJ}$ (panel B).  Red lines display
the range of value observed for the current $D_3$ population, from 0.0031 to 
0.0035.  The vertical green lines display the value of minimum 
estimated escape velocity from the Koronis parent body. The horizontal blue
lines in panel B show values of $\cos{(\omega +f)} = \frac{\sqrt{2}}{2}$.}
\label{fig: f_wf_VEJ}
\end{figure*}
We first analyzed how the current $e$ distribution of the $a < 2.88$~au 
population depends on $f$.  For this
purpose, we generated a grid of 31 by 13 fictitious Koronis families
with values of $V_{EJ}$, the parameter determining the spread of the initial 
ejection velocity field (see Eq. 5 in \citet{Carruba_2015c}, going from
50 to 200~m/s, and $f$ from 0 to 360$^{\circ}$.  We then computed the standard 
deviation of these fictitious families, and compared their values with the 
current one for the $D_3$ population  (0.0190), with an assumed 5\% error 
(at this stage of our analysis, we neglected the effect of dynamical evolution 
in proper $e$, see Fig.~\ref{fig: VEJ_F_E} for a qualitative understanding
of the dependence of ${\sigma}_{e}$ on $V_{EJ}$ and $f$).  The minimum 
possible value of $V_{EJ}$ is 170~m/s, which is quite higher than the 
estimated escape velocity from the 
Koronis parent body (130.6~m/s, assuming that the diameter of the parent body
was 122 km, the minimum value in the literature, \citet{Nesvorny_2015}). 
This may suggest that the current distribution in proper $e$ of the 
$a < 2.88$~au population might be significantly affected by dynamical evolution.
The situation is different for the $a < 2.88$~au $\sin{(i)}$.  Again,
we generated fictitious Koronis families with different values of $f$ 
and $(f+\omega)$, and of $V_{EJ}$, compute their standard deviation,
and compared with the current one for the $D_3$ population  (0.0033), 
with an assumed 5\% error.  Fig.~\ref{fig: f_wf_VEJ} displays our results in 
the $(V_{EJ},f)$ (panel A) and $(V_{EJ},\cos{(\omega+f)})$ (panel B) planes.  
The vertical green line displays the estimated escape velocity from the 
Koronis parent body.  In the first case we assumed that 
$(\omega+f)= 45^{\circ}$, while in the second we used $f=180^{\circ}$. 
  
Our results show that the $\sin{(i)}$ distribution does not depend 
significantly on $f$, as expected from the analysis of the denominator of 
Eq.~\ref{eq: gauss_3}, and the fact that the mean eccentricity of Koronis 
members is small (of the order of 0.05).  Essentially values of 
$V_{EJ}$ in a strip from 75 to 100 m/s would all produce families 
consistent with the current distribution in $\sin{(i)}$ regardless of 
the original value of $f$.  The situation is different for $\omega+f$.  
Using $f = 180^{\circ}$, the value of 
$f$ that provided the best results in the previous analysis (
all values of $f$ are of course admissable, according to 
Eq.~\ref{eq: gauss_3}, here we just picked the one that 
provided the most optimal result in our previous analysis), we obtain values 
of the standard deviation in $\sin{(i)}$ as a function of $V_{EJ}$ and 
$\omega+f$ (Fig.~\ref{fig: f_wf_VEJ}, panel B).  The minimum value
of $\cos(\omega +f) V_{EJ}$ is of the order of 60~m/s.   If one assumes
that $V_{EJ}$ does not much exceed the escape velocity, since this
would imply very energetic impacts, which are quite rare 
\citep{Bottke_2015}, then values of 
$\cos{(\omega+f)}$ in the range from -0.2 to 0.2 can be excluded.  
According to this analysis, values of $\omega+f$ from $78.5^{\circ}$ to 
$101.5^{\circ}$ (and the analogous negative range) should therefore be unlikely.

\begin{figure}
\centering
\centering \includegraphics [width=0.47\textwidth]{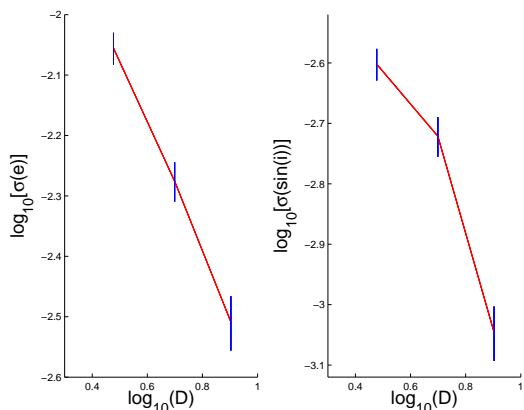}

\caption{Log-log plots of the dependence of the standard deviation
of the distribution in $e$ and $\sin{(i)}$ of the Koronis SDDS family 
members, as a function of diameter, when corrections caused by the
dynamical evolution are accounted for.  Vertical blue lines display
the nominal errors, assumed to be inversely proportional to the square 
root of the number of objects in each size-bin.}
\label{fig: Dsigma_e_sini_corr}
\end{figure}

What values of the $V_{EJ}$ parameter would we expect for the initial
ejection velocity field of the Koronis family, before dynamical effects 
occurred?  In Sect.\ref{sec: loc_dyn}
we computed changes in $\sin{(i)}$ for simulated members of the 
Koronis family, for three different size ranges.  If we assume that 
changes in $\sin{(i)}$ caused by dynamical processes and by the initial 
velocity distribution can be summed as two independent
random variables, which should be the case for families, such as Koronis,
not affected by powerful secular resonances such as the ${nu}_6$, $z_1$ or 
$z_2$ resonances, then the initial standard deviation of $\sin{(i)}$ 
associated with the original ejection velocity field, ${\sigma}_{sin(i)}$, can 
be obtained from the relationship: 
\begin{equation}
{\sigma}_{sin(i)}=\sqrt{{{\sigma}_{cur}}^2-{{\sigma}_{dyn}}^2},
\label{eq: sigma_in}
\end{equation}
\noindent where ${\sigma}_{cur}$ is standard deviation of the current 
distribution of either $e$ or $\sin{(i)}$ values, and ${\sigma}_{dyn}$
is the standard deviation of changes caused by dynamics.   We computed
the standard deviations of $e$ and $\sin{(i)}$ corrected 
for the effects of dynamical evolution using Eq.~\ref{eq: sigma_in}
and data from Table~\ref{table: dedi_std_fict}, as a function
of asteroid diameters.  If we use the data
on the diffusion caused in the area with $a < 2.88$~au, the distributions 
still follow a power-law of the form $\sigma = C (\frac{1}{D})^{\alpha}$,
with $\alpha=1.02$ and $0.95$, with the same errors discussed 
in Sect~\ref{sec: fam_ide}.  To within a 9\% error, this law 
is compatible with an inverse relationship of the 
form $\sigma = C (\frac{1}{D})$, that also applies to $v_W$ 
through Eq.~\ref{eq: gauss_3}.   If we use data on the dispersion
computed for the whole family, and obtain ${\sigma}_{cur}$ using 
Eq.~\ref{eq: sigma_in}, then the values of ${\sigma}_{cur}$, 
shown in Fig.~\ref{fig: Dsigma_e_sini_corr}, are lower. However,
$\alpha=1.06\pm 0.14$ and $1.03\pm 0.13$ are still compatible
with an inverse relationship.

\begin{figure*}[!htp]
\centering
  \begin{minipage}[c]{0.47\textwidth}
    \centering \includegraphics[width=2.8in]{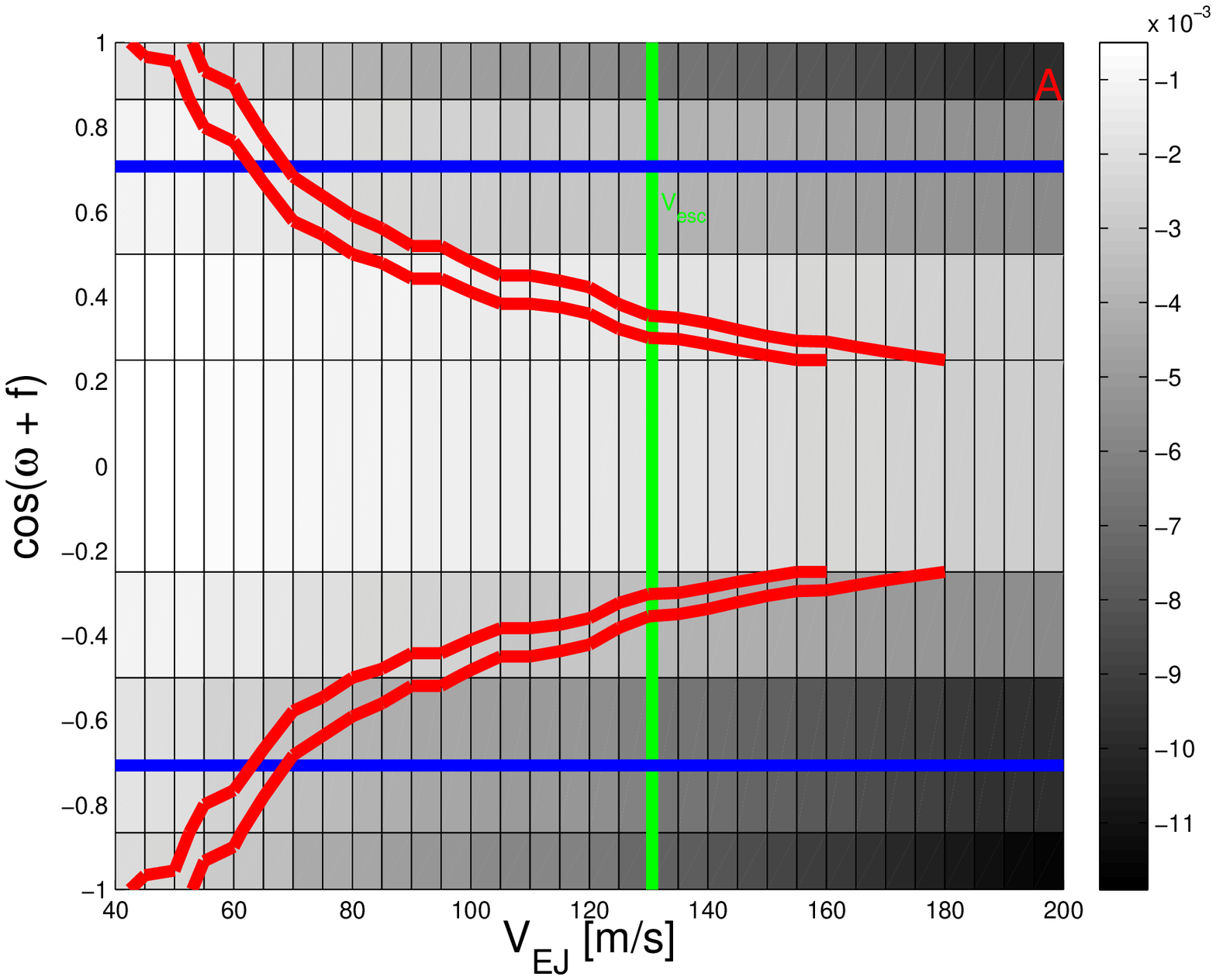}
  \end{minipage}%
  \begin{minipage}[c]{0.47\textwidth}
    \centering \includegraphics[width=2.8in]{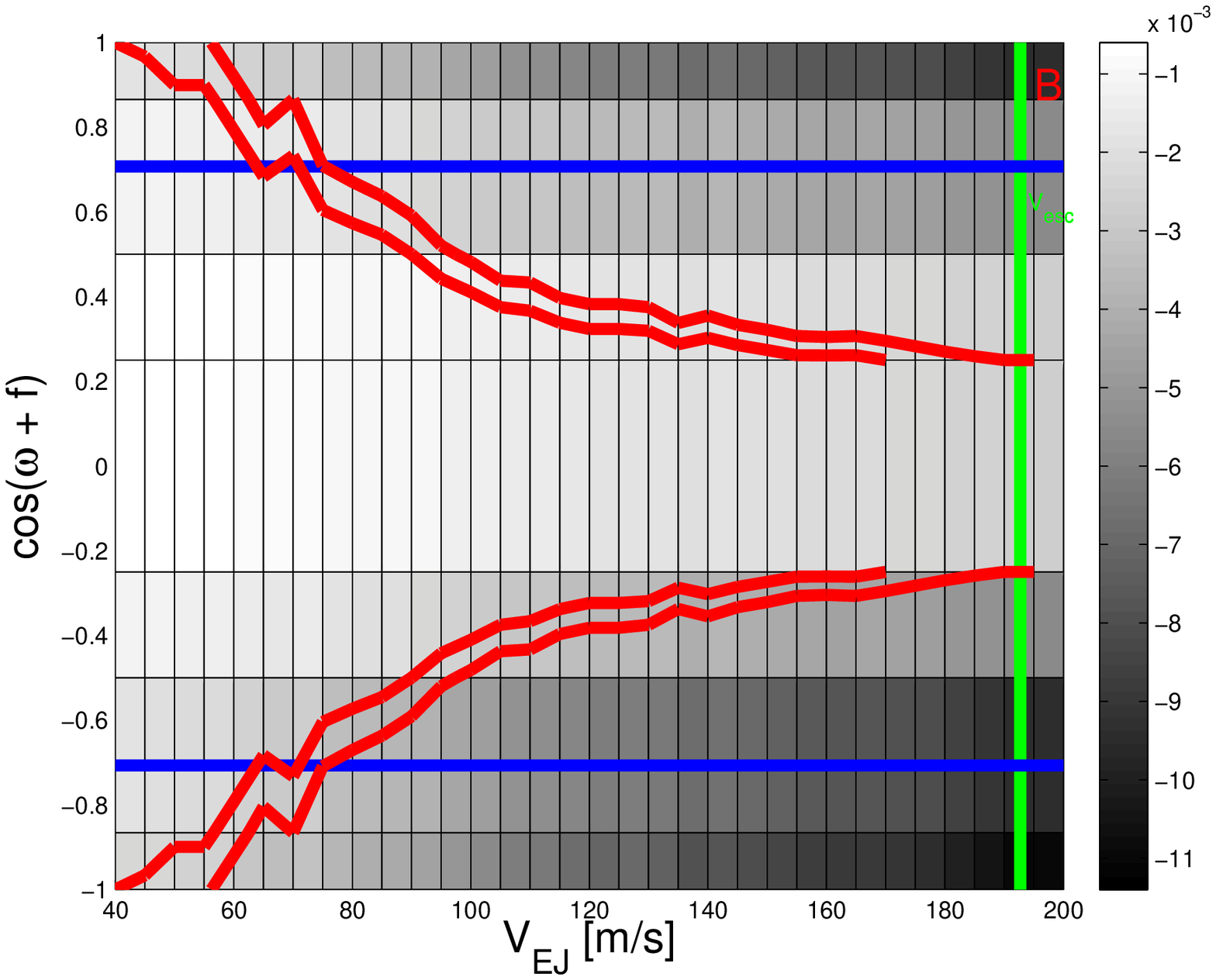}
  \end{minipage}
\caption{Values of the standard deviation in $\sin{(i)}$ of simulated Koronis 
family $D_3$ populations and $D_{PB} = 122$ km (Panel A), and $D_{PB} = 180$ km 
(Panel B). The vertical green lines display the values of escape velocity, 
while blue lines are associated with $\omega +f = 45^{\circ}$.  The 
red lines have the same meaning as in Fig.~\ref{fig: f_wf_VEJ}.}
\label{fig: VEJ_wf}
\end{figure*}

Assuming that that is true, what value of $V_{EJ}$ in Eq.~\ref{eq: std_veJ}
could best describe the initial spread in $\sin(i)$ (and possibly $e$)?  
We created various fictitious families for the three Koronis size intervals 
in $D$ previously studied ($6 < D < 10, 4 < D < 6, 2 < D < 4$~km), and 
two values of the diameter of the Koronis parent
body, one obtained extrapolating the current observed SFD down to 
zero km ($D_{PB}$=122 km), and the second by using \citet{Durda_2007} 
approach ($D_{PB}$=180 km, see also \citet{Nesvorny_2015}).  Since
the SFD's of old families are usually depleted by collisional and
dynamical evolution, \citet{Durda_2007} approach that use only bodies
with $D > 10$ km is often preferred.  We assume 
that the initial ejection speeds were 
isotropic, and we used different values of $\cos{(\omega+f)})$ in the 
range from -1 to 1, and $V_{EJ}$ from 40 to 200 m/s.  We did not sample 
the interval of $\cos{(\omega+f)})$ between -0.2 and 0.2, since this
was excluded by our previous analysis. For the values
of the current spread, we used those corrected for the effects of
the whole Koronis family dynamics.

Our results for the $D_3$ population are displayed in 
Fig.~\ref{fig: VEJ_wf} (for the sake of brevity, we do not show
the results for the $D_5$ and $D_8$ size populations).  With the assumption 
of isotropic initial speeds, minimum values of $V_{EJ}$ are in the range 
from 40 to 55 m/s.  Distributions are in good
agreement, and compatible with what obtained by Yarko-Yorp methods
for the distribution in proper $a$ (70 m/s, Carruba et al. 2015b).

Finally, Fig.~\ref{fig: VEJ_F_E} shows a contour plot of standard deviations 
in eccentricity for simulated Koronis family $D_3$ population, with
$D_{PB} = 122$ km (for the sake of brevity, we do not show the results for 
$D_{PB} = 180$).  If we account for the effects of dynamics on the whole
Koronis family, the new minimum value of $V_{EJ}$, 80~m/s is now lower than 
the minimum escape velocity from Koronis parent body, and in agreement
with results from the $\sin{(i)}$ distribution, assuming that 
$-0.6 < \cos(\omega +f) < 0.6$.   

The large correction from dynamical effects,
not observed for the $a < 2.88$ au population, may be caused by the fact that
our Koronis simulated sample was limited (502 particles) and/or not able to 
fully reproduce the effect of particles reorienting their spin axis and 
inverting the direction of migration from higher to lower $a$.
Perhaps this may suggest that  the stochastic YORP effect may not be
that ``stochastic'', meaning that longer timescales for choosing new
asteroid shape models (and consequently, more reorientation events) could
be closer to what happens in nature.  A static YORP effect is actually needed
to explain the ecliptic latitude distribution of main belt asteroids
\citep{Hanus_2013} or the Slivan states of some Koronis family members
\citep{Vokrouhlicky_2003}.
Alternatively, if we only consider the results of our dynamical simulation
for the $a < 2.88$ au population, the initial ejection velocity field could 
have been rather unisotropical, with lower values of $\delta v_W$ and 
larger values of $\delta v_r$ and  $\delta v_t$.  Further study is needed to
clarify this issue.

\begin{figure}
\centering
\centering \includegraphics [width=0.47\textwidth]{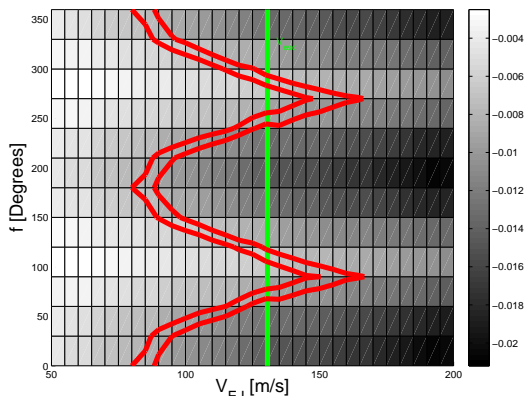}

\caption{Values of the standard deviation in $e$ of simulated Koronis 
family $D_3$ populations and $D_{PB} = 122$ km. Red and green lines
have the same meaning as in Fig.~\ref{fig: VEJ_wf}.}
\label{fig: VEJ_F_E}
\end{figure}

\section{Conclusions}
\label{sec: conc}
The main results of this work can be summarized as follows:
\begin{itemize}

\item We identified the Koronis family in the domain of proper elements
and used SDSS-MOC4 and WISE data to identify members of the Koronis halo,
so as to obtain better estimates of the current spread in proper
$e$ and $\sin(i)$ of the family.  It has been assumed that the spread of the 
original ejection velocity field (and therefore of $e$ and $\sin(i)$, 
\citet{Carruba_2015c}) should actually be inversely proportional to $D$ 
\citep{Vokrouhlicky_2006}. Here we show that this is actually 
the case for the Koronis family.  
\item We studied how the local dynamics may have affected the original
distribution in proper $e$ and $\sin(i)$ of the Koronis family by
obtaining dynamical maps of the Koronis orbital region, and by performing
numerical integrations of a fictitious family with the $SYSYCE$ integrator 
of \citet{Carruba_2015a}.  Local dynamics for the whole family 
affects more eccentricities than inclination: about all of the current
spread in eccentricity could be caused by dynamical effects.  Conversely,
up to 50\% of the current spread in inclination can be original.
\item We estimated the original dispersion of the Koronis family in
proper eccentricity and inclination, assuming that changes caused by 
dynamical processes and by the initial velocity distribution can be 
summed as two independent random variables.  Estimated values of the 
$V_{EJ}$ parameter describing the initial spread of the Koronis family 
should be of the order of 80 m/s, and compatible with what previously 
found with Yarko-YORP methods to fit the Koronis family semi-major axis 
distribution (70 m/s).
\end{itemize}
Overall, for the first time we obtained estimates of the $V_{EJ}$ parameter
describing the initial spread of the Koronis family using the {\it inclination}
distribution, rather than the semi-major axis one, as previously attempted.
Our results suggests that i) the initial spread in ejection velocity
might indeed be inversely dependent on the asteroid sizes, as previously
assumed by other authors \citep{Vokrouhlicky_2006}, and ii) 
values of $V_{EJ}$ could be compatible with what previously obtained by 
using Yarko-Yorp methods to fit the semi-major axis distribution of the
Koronis family.  Extending this analysis to other families identified
in \citet{Carruba_2015c} as good candidates for not being too much affected
by dynamical evolution, remains a challenge for possible future research.
\section*{Acknowledgments}
We are grateful to the reviewers of this paper, Drs. 
Bojan Novakovi\'{c} and Miroslav Bro\v{z}, for comments and suggestions
that greatly improved the quality of this work.  We thank  
Dr. William F. Bottke for allowing us to use the results of his dynamical
simulation of the Koronis family from the \citet{Bottke_2001} paper.
This paper was written while the first author was a visiting scientist
at the Southwest Research Institute (SWRI) in Boulder, CO, USA.
We would like to thank the S\~{a}o Paulo State Science Foundation 
(FAPESP) that supported this work via the grants 14/24071-7 and 
13/15357-1. D.N.'s work on this project was supported by NASA's Solar 
System Workings program.  This publication makes use of data products from 
the Wide-field Infrared Survey Explorer, which is a joint project of the 
University of California, Los Angeles, and the Jet Propulsion 
Laboratory/California Institute of Technology, funded by the National 
Aeronautics and Space Administration.  This publication also makes use of 
data products from NEOWISE, which is a project of the Jet Propulsion 
Laboratory/California Institute of Technology, funded by the Planetary 
Science Division of the National Aeronautics and Space Administration.

\end{document}